\pdfoutput=1
\documentclass[namedreferences]{SolarPhysics}
\usepackage[optionalrh]{spr-sola-addons} 
\usepackage{graphicx}        
\usepackage{gensymb}        
\usepackage{color}           
\usepackage{url}             
\usepackage[pdfborder={0 0 0 },urlcolor=blue]{hyperref}
\ifx \doiurl \undefined \def \doiurl#1{\href{http://dx.doi.org/#1}{\url{#1}}}\fi
\ifx \adsurl \undefined \def \adsurl#1{\href{http://adsabs.harvard.edu/abs/#1}{\url{#1}}}\fi

\newcommand{\aap}{    {\it Astron. Astrophys.}}

\newcommand{\apj}{    {\it Astrophys. J.}}
\newcommand{\apjl}{   {\it Astrophys. J. Lett.}}

\newcommand{\mnras}{  {\it Mon. Not. Roy. Astron. Soc.}}

\newcommand{\pasj}{   {\it Pub. Astron. Soc. Japan}}

\newcommand{\solphys}{{\it Solar Phys.}}

\begin{document}

\begin{article}

\begin{opening}

\title{Properties of High-Frequency Wave Power Halos around Active Regions: An Analysis of Multi-Height Data from HMI and AIA 
Onboard SDO.\\}

\author{S.P.~\surname{Rajaguru}$^{1}$\sep
        S.~\surname{Couvidat}$^{2}$\sep
        Xudong~\surname{Sun}$^{2}$\sep
        K.~\surname{Hayashi}$^{2}$\sep
        H.~\surname{Schunker}$^{3}$
       }
\runningauthor{S.P. Rajaguru {\it et al.}}
\runningtitle{Acoustic Halos}

   \institute{$^{1}$ Indian Institute of Astrophysics, Bangalore - 34, India 
                     \href{mailto:rajaguru@iiap.res.in}{rajaguru@iiap.res.in}\\
              $^{2}$ Hansen Experimental Physics Lab, Stanford University, Stanford CA, USA 
                     \href{mailto:couvidat@stanford.edu}{couvidat@stanford.edu} \\
              $^{3}$ Max-Planck Institute for Solar System Research, Katlenburg-Lindau, Germany
                     \href{mailto:schunker@mps.mpg.de}{schunker@mps.mpg.de} \\
             }

\begin{abstract}
We study properties of waves of frequencies above the photospheric acoustic cut-off of $\approx$ 5.3 mHz,
around four active regions, through spatial maps of their power estimated using data from the \textit{Helioseismic and 
Magnetic Imager} (HMI) and \textit{Atmospheric Imaging Assembly} (AIA) onboard the \textit{Solar Dynamics Observatory} (SDO). 
The wavelength channels 1600 \AA~ and 1700 \AA~ from AIA are now known to capture clear oscillation signals 
due to helioseismic {\em p}-modes as well as waves propagating up through to the chromosphere. Here we study in detail, 
in comparison with HMI Doppler data, properties of the power maps, especially the so called ``acoustic halos''
seen around active regions, as a function of wave frequencies, inclination, and strength of magnetic field
(derived from the vector-field observations by HMI) and observation height. We infer possible signatures
of (magneto)acoustic wave refraction from the observation-height dependent changes, and hence due to
changing magnetic strength and geometry, in the dependences of power maps on the photospheric magnetic quantities. 
We discuss the implications for theories of {\it p}-mode absorption and mode conversions by the magnetic field.
\end{abstract}
\end{opening}

\section{Introduction}
\label{sec:intro}

Enhanced power of high-frequency waves surrounding strong-magnetic-field structures such as sunspots and
plages is one of several intriguing wave-dynamical phenomena observed in the solar atmosphere.
This excess power known as ``acoustic halo'', first observed in the early 1990's at photospheric
\cite{brownetal92} as well as chromospheric \cite{braunetal92,tonerlabonte93} heights, is at frequencies
above the photospheric acoustic cut-off of $\approx$ 5.3 mHz, in the range of 5.5 -- 7 mHz, and over
regions of weak to intermediate strength (50 -- 250 G) photospheric magnetic field.
A good number of observational studies \cite{hindmanandbrown98, thomasandstanchfield00,
jainandhaber02, finsterleetal04, morettietal07, nagashimaetal07} since then have brought out additional features,
and we refer the reader to \inlinecite{khomenkoandcollados09} for a succint summary of them as known prior to 2009.
On the theoretical side, no single model describing all of the observed features has been achieved yet, although there
have been several focussed efforts \cite{kuridzeetal08,hanasoge08,shelyagetal09,hanasoge09,khomenkoandcollados09}.
However, a large number of studies centered around modeling acoustic wave -- magnetic-field interactions
over heights from the photosphere to chromosphere, with relevance to high-frequency power excess observed around 
sunspots, have been carried out \cite{rosenthaletal02,bogdanetal03,bogdanandjudge06,cally06,schunkerandcally06,
khomenkoandcollados06,khomenkoandcollados08,jacoutotetal08,khomenkoetal09,vigeeshetal09,khomenkoandcally12}.
A central theme of all the above theoretical studies, except that of \inlinecite{jacoutotetal08}, 
has been the conversion of acoustic wave modes (from below the photosphere) into magnetoacoustic wave modes 
(the fast and slow waves) at the magnetic canopy defined by the plasma $\beta$=1 layer. Enhanced acoustic emission by 
magnetically modified convection over weak and intermediate field regions, suggested as one possible mechanism by 
\inlinecite{brownetal92} and further advocated
by \inlinecite{jainandhaber02}, was found to be viable, through 3D numerical simulations of magneto-convection, by
\inlinecite{jacoutotetal08}.
It should perhaps be noted here that Hindman and Brown (1998) suggested some form of 
field-aligned incompressible wave motions as agents for excess power over magnetic regions, implied 
by their finding, from SOHO/MDI observations, of visibility of these halos in photospheric Doppler velocities but not
in continuum intensities. This suggestion, however, seems to contradict chromospheric-intensity observations 
\cite{braunetal92, morettietal07}, which show that the halos at these heights, in terms of their dependence on
magnetic field strength as well as frequency behaviour, are clearly related to the photospheric halos; Hindman and Brown's
reasoning that the chromospheric Ca K intensities observed by \inlinecite{braunetal92} have large cross-talk from
Doppler shifts are however contradicted by the simultaneous velocity and intensity observations
made by \inlinecite{morettietal07}. 

In a recent study, \inlinecite{schunkerandbraun11} have brought out a few new properties, 
{\it viz.} i) the largest excess power in halos is at
horizontal magnetic field locations, in particular, at locations between opposite-polarity regions, ii) the
larger the magnetic-field strength the higher the frequency of peak power, and iii) the modal ridges over
halo regions exhibit a shift towards higher wavenumbers at constant frequencies. Though none of the proposed theoretical
explanations or mechanisms causing the power halos are able to match all of the observed properties, and hence provide
an acceptable theory, the mechanism based on MHD fast-mode refraction in canopy-like structure of strong expanding
magnetic field studied by Khomenko and Collados (2009) appears to match some major observed features. This theory
also predicts certain other observable features that we will address here. 

From what has been learned so far, from observations as well as theoretical studies,
it is clear that transport and conversion of energy between magneto-acoustic wave modes, which are driven
by acoustic waves and convection from below the photosphere and mediated by the structured magnetic
field in the overlying atmosphere, provide a plausible approach for identifying the exact mechanism. A crucial disgnostic of such
wave processes requires probing several heights in the atmosphere simultaneously with magnetic-field information.
The instruments HMI and AIA onboard SDO, with photospheric Doppler and vector magnetic field information from 
the former and the upper photospheric and lower chromospheric UV emissions in the 1700 \AA~  and 1600 \AA~ wavelength 
channels of the latter, provide some interesting possiblities for such studies. We exploit this opportunity, and make a detailed
analysis of high-frequency power halos around four different active regions, over at least five different
heights from the photosphere to chromosphere.

\begin{figure}
\centerline{\includegraphics[width=1.5\textwidth,height=0.8\textheight,clip=]{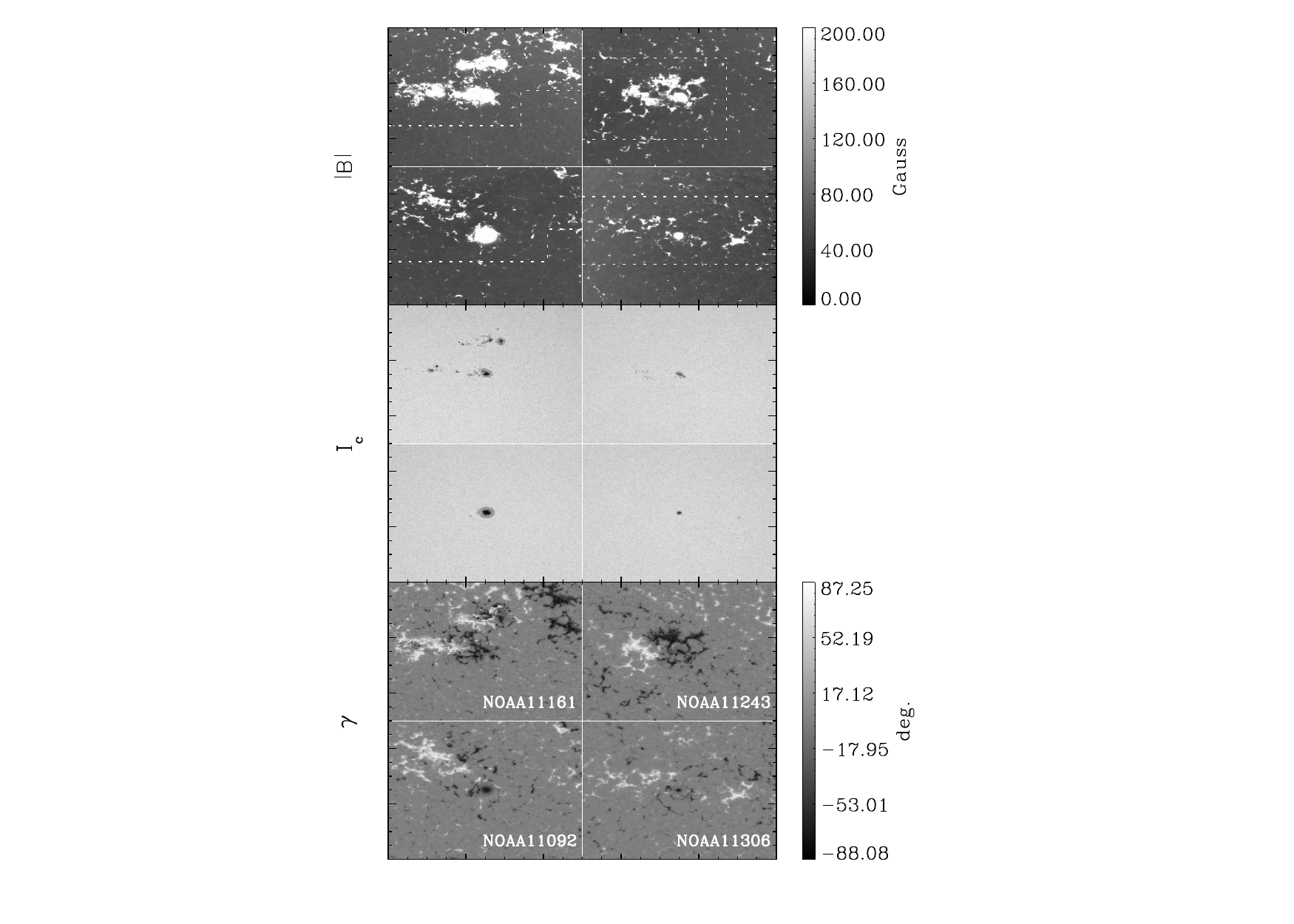} }
\caption{Average total magnetic field [$B$] (top panel, pixels above 200 G are saturated in the grey scale),
a snapshot of continuum intensity [$I_{\rm{c}}$] (middle) and average magnetic-field inclination [$\gamma$] (bottom) of the four
active regions studied. Averages of $B$ and $\gamma$ shown here are over the length of time (14 hours) used for power map estimation.
Areas covered between white dashed lines and boundary axes in the top panel, in each region, are the largely non-magnetic ones used
for estimation of quiet-Sun power of waves studied here. Each active region here covers a square area of 373 $\times$ 373 Mm$^{2}$. }
\label{fig1}
\end{figure}
 
\section{Data and Analysis Method}
\label{sec:data}
We use data from \textit{Helioseismic and Magnetic Imager} (HMI: Scherrer {\it et al.} 2012) and \textit{Atmospheric Imaging Assembly}
(AIA: Lemen {\it et al.} 2012) onboard the \textit{Solar Dynamics Observatory} (SDO): photospheric Doppler velocity [$v$], 
continuum intensity [$I_{\rm{c}}$],
line-core intensity [$I_{\rm{co}}$], and dis-ambiguated vector magnetic field [$B_{\rm{x}}, B_{\rm{y}},$ and $B_{\rm{z}}$] derived from
HMI observations, and chromospheric UV emissions observed by AIA in the wavelength channels 1700 \AA~ and 1600 \AA, which
we denote as $I_{\rm{uv1}}$ and $I_{\rm{uv2}}$, respectively. The intensities $I_{\rm{uv1}}$ and $I_{\rm{uv2}}$ are now known to capture clear
oscillation signals due to helioseismic p modes as well as propagating waves in the atmosphere \cite{howeetal10,hilletal11}. 
The photospheric observations by HMI are in the form of filtergram images captured from across the magnetically
sensitive line Fe {\sc i} 6173.34 \AA~ using two different cameras: the Doppler camera uses six images each of the left- and
right-circular polarization components (Stokes $I$+$V$, and $I$-$V$) to measure $v$, $I_{\rm{c}}$, the line depth
[$I_{\rm{ld}}$] and line width with a
cadence of 45 seconds \cite{scherreretal12}, while the second vector field camera records the full Stokes vector [$I, Q, U, V$] 
in six combinations over six wavelengths (i.e., a total of 36 filtergrams) with a cadence of 135 seconds \cite{hoeksemaetal12}. 
It should perhaps be pointed out 
that $I_{\rm{c}}$, $I_{\rm{co}}$, and $v$ from HMI form at three different heights spread over 
$z$ = 0 -- 300 km above the continuum optical depth $\tau_{\rm{c}}$ = 1 ($z$ = 0 km) level \cite{nortonetal06}: 
$I_{\rm{c}}$ is from about $z$ = 0 km, $v$ corresponds to an average height
of about $z$ = 140 km \cite{flecketal11}, and the line core intensity calculated as $I_{\rm{co}} = I_{\rm{c}} - I_{\rm{ld}}$ corresponds
to the top layer, at about $z$ = 280 -- 300 km, of the line formation region \cite{nortonetal06,flecketal11}. 
In addition, exploiting the availability of individual
filtergram (level 1) images from HMI, we derive a Doppler velocity [$v_{50}$] from only the outer pair of filtergrams 
[$I_{0}$ and $I_{5}$], which are measured at +172 m\AA~ and -172 m\AA~, respectively, from the rest frame line center wavelength
of 6173.34 \AA: $v_{50} = k_{50}(I_{5} - I_{0})/(I_{5}+I_{0})$, where the calibration
constant $k_{50}$ is derived from the spacecraft velocity, {\sf{OBSVR}}, known to very good accuracy. Filtergrams $I_{0}$ and $I_{5}$
sample the line wings and hence a height corresponding to the lower end within the line formation
region \cite{nortonetal06}, {\em i.e.} at $z$ = 20 km. We should note here that the positions of $I_{0}$ and $I_{5}$, and
hence the height level that they sample, depend on Zeeman broadening and the line shifts due to the spacecraft velocity:
in fields stronger than about 1000 G and when {\sf{OBSVR}} exceeds about 2 km s$^{-1}$, these wing filtergrams do not sample the
blue and red wings symmetrically and this not only invalidates the above height identification, but also corrupts the 
Doppler-velocity estimate. Since we focus only on the power excess, which is seen over weak and intermediate field strengths not
exceeding about 800 G (see Figures 5 -- 10), use of $v_{50}$ in our analysis is not subject to above uncertainties (on 
height of formation and Doppler amplitudes). The AIA 1700 \AA~ and 1600 \AA~ intensities
[$I_{\rm{uv1}}$ and $I_{\rm{uv2}}$], form at average heights of 360 km and 430 km \cite{fossumandcarlsson05}, respectively. Thus, we have 
velocities and intensities from at least five different heights ranging from $z=$ 0 to 430 km.

\begin{figure}    
\centerline{\hspace*{0.16\textwidth}
               \includegraphics[width=1.1\textwidth,height=0.6\textheight,clip=]{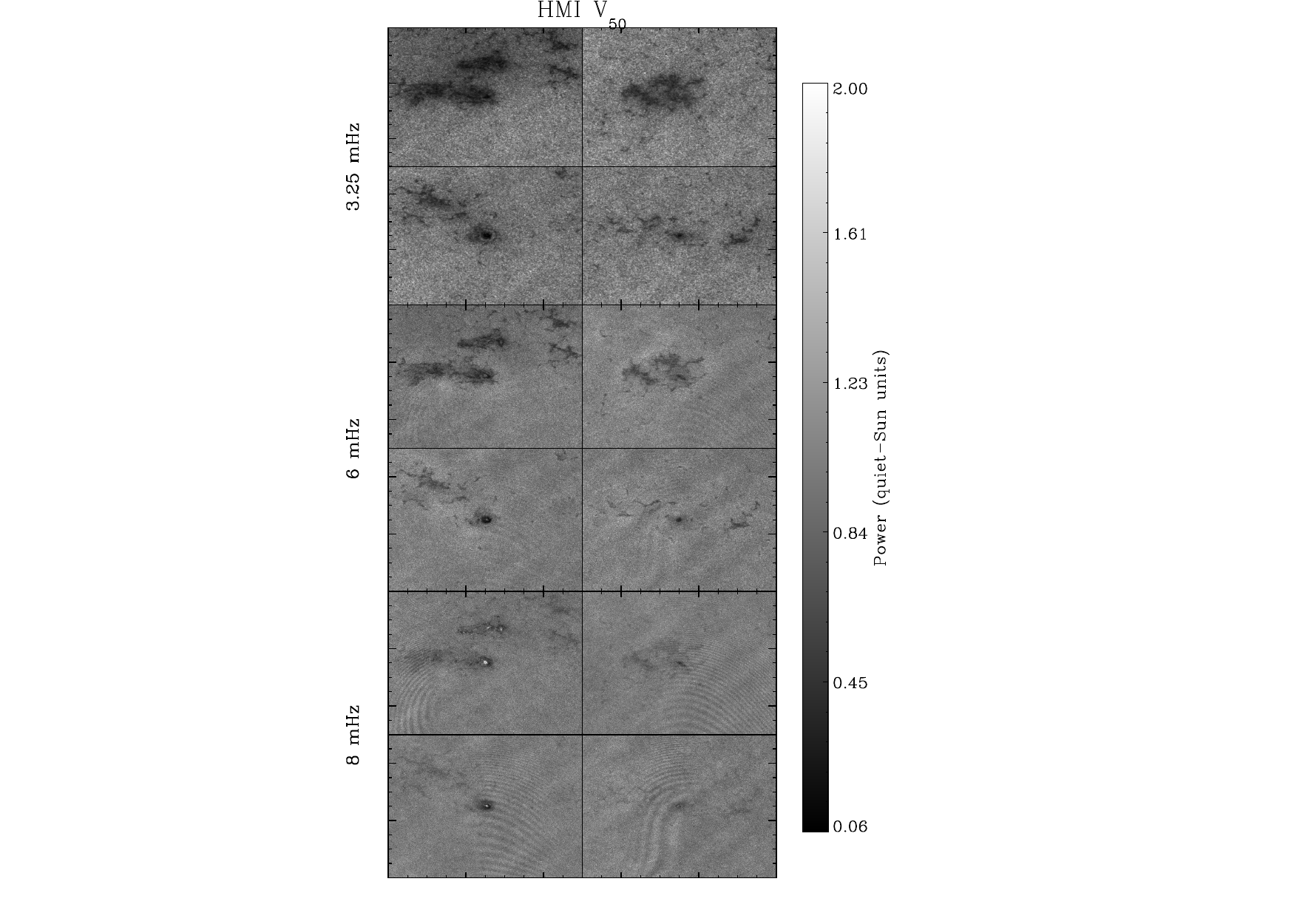}
               \hspace*{-0.65\textwidth}
               \includegraphics[width=1.1\textwidth,height=0.6\textheight,clip=]{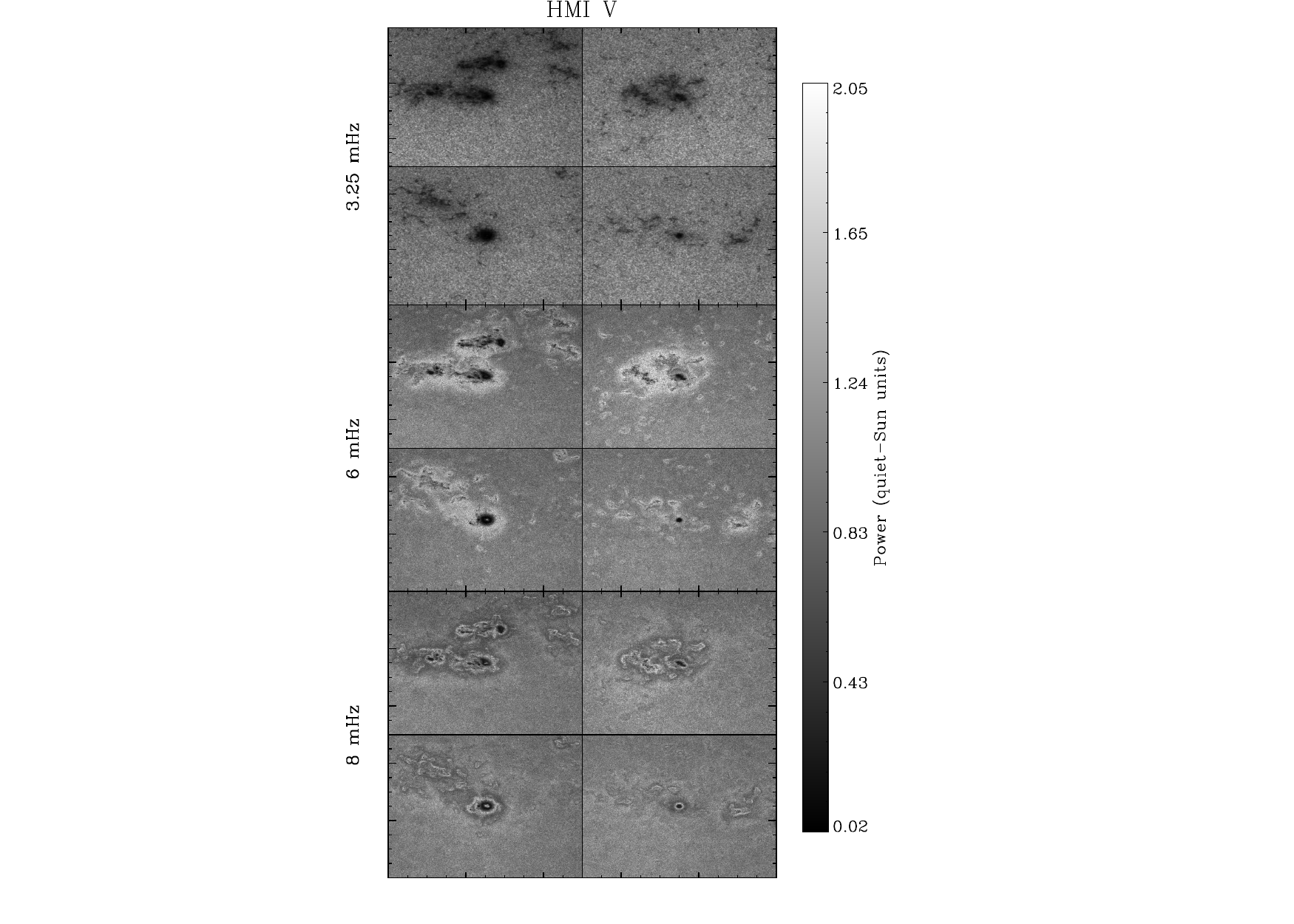}
              }
     \vspace{-0.986\textwidth}   
     \centerline{\Large \bf     
      \hspace{0.15\textwidth}  \color{black}{(a)}
      \hspace{0.39\textwidth}  \color{black}{(b)}
         \hfill}
     \vspace{0.93\textwidth}    
\caption{Normalized, with respect to quiet Sun, power over the four active regions (arranged as in Figure 1), 
each covering a square area of 373 $\times$ 373 Mm$^{2}$, at three representative frequencies: 3.25, 6, and 8 mHz (top, middle, and 
bottom panels, respectively). Left part, (a), of the figure is obtained from $v_{50}$ and the right part, (b), is from $v$.}
\label{fig2}
\end{figure}
Our chosen target regions of study are the four active regions, NOAA 11092, 11161, 11243, and 11306, each of spatial size covered in 
512 $\times$ 512 pixels, with a sampling rate of 0.06 degrees (heliographic) per pixel, and tracked at the Carrington rotation rate for 
about 14 hours each during their 
central meridian passage dates of 3 August 2010, 11 February 2011, 3 July 2011, and 2 October 2011, respectively. 
Both the HMI and AIA data cubes for the above regions are temporally and spatially aligned through the tracking and remapping 
(Postel) routines of the Stanford SDO data pipeline systems. The vector magnetic field over the four regions are determined using 
the HMI vector-field pipeline, which does the following: Stokes parameters derived from filtergrams observed over a 12-minute 
interval are inverted using a Milne--Eddington based algorithm, the {\em Very Fast Inversion of the Stokes Vector} (VFISV: 
Borrero {\em et al.} 2011); the 180\degree azimuthal ambiguity in transverse field is resolved by an improved version of the {\em minimum energy 
algorithm }\cite{metcalf94,metcalfetal06,lekaetal09}. A detailed description of the production and relevant characteristics, 
and outstanding questions regarding HMI vector-field data reduction are also described by \inlinecite{hoeksemaetal12}. 
The vector field maps over the target regions are tracked and remapped the same way as for the other data.

From the disambiguated vector field maps [$B_{\rm{x}}, B_{\rm{y}}$, $B_{\rm{z}}$] we determine the field inclination [$\gamma$] defined as 
$\tan(\gamma) = B_{\rm{z}}/B_{\rm{h}}$, where $B_{\rm{h}}=\sqrt{B_{\rm{x}}^{2}+B_{\rm{y}}^{2}}$, 
similarly to Schunker and Braun (2011). Hence, 
$\gamma$ ranges from $-90 \degree$ to  $+90 \degree$ with $\gamma$ = 0\degree when the field is purely horizontal, 
$\gamma < 0 \degree$ when $B_{\rm{z}} < 0$ and $\gamma > 0\degree$ when $B_{\rm{z}} > 0 $. Figure 1 displays images of $I_{\rm{c}}$, 
absolute total field strength $B=\sqrt{B_{\rm{x}}^{2}+B_{\rm{y}}^{2}+B_{\rm{z}}^{2}}$, and $\gamma$; the latter two are averages over 
the period of observation used in this work. 
\begin{figure}    
\centerline{\hspace*{0.16\textwidth}
               \includegraphics[width=1.1\textwidth,height=0.6\textheight,clip=]{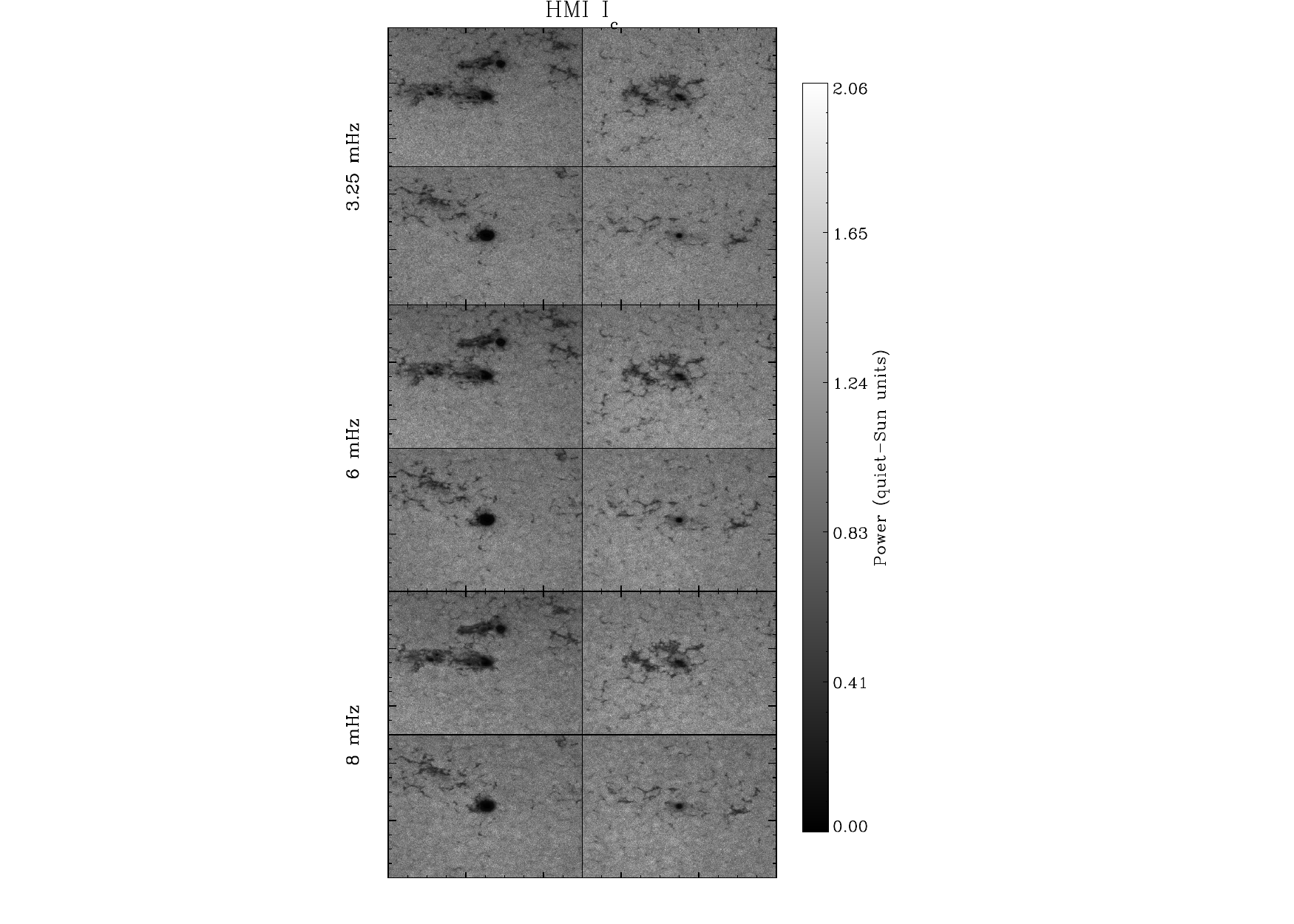}
               \hspace*{-0.65\textwidth}
               \includegraphics[width=1.1\textwidth,height=0.6\textheight,clip=]{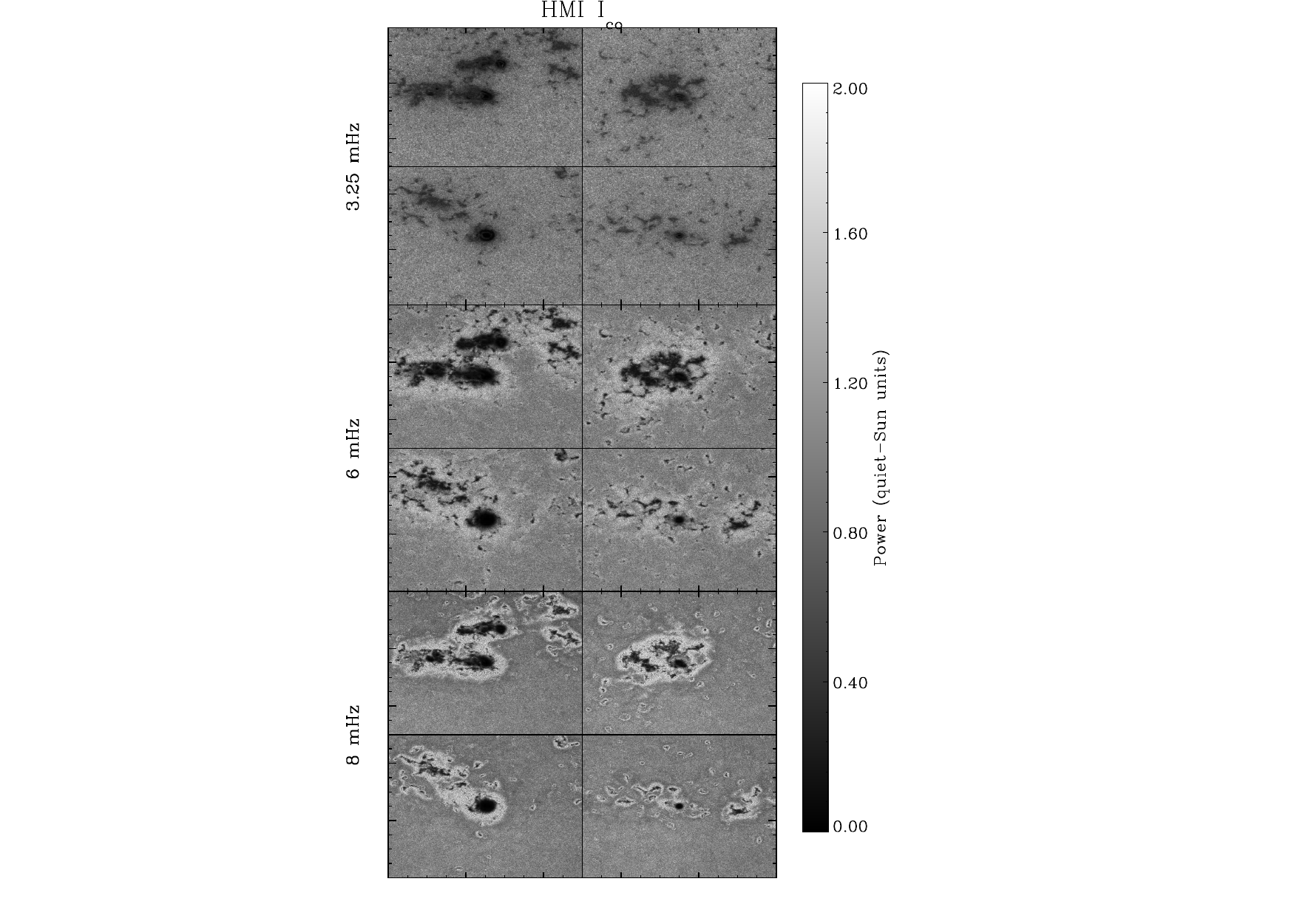}
              }
     \vspace{-0.986\textwidth}   
     \centerline{\Large \bf     
      \hspace{0.15 \textwidth}  \color{black}{(a)}
      \hspace{0.39\textwidth}  \color{black}{(b)}
         \hfill}
     \vspace{0.93\textwidth}    
\caption{Normalized power over the four active regions (arranged as in Figure 1)
at three representative frequencies, 3.25, 6, and 8 mHz (top, middle, and bottom panels, respectively) obtained from
observables HMI $I_{\rm{c}}$ [(a)] and $I_{\rm{co}}$ [(b)]. Each active region covers a square area of 373 $\times$ 373 Mm$^{2}$.}
\label{fig3}
\end{figure}

Using the 14-hour long time series of images prepared as above, we calculate maps of power summed over 1-mHz band of 
frequencies centered every 0.25 mHz in the frequency range of 2 -- 10.75 mHz, and normalise them with average power estimated
over a quiet-Sun patch identified on each date (i.e. on each target region, see Figure 1) separately.  Figures 2 -- 4 display 
the power maps obtained from the different observables, which correspond to different heights in the atmosphere as explained above.
In the following sections we present results of our analyses of these power maps as a function of $B$ and $\gamma$.
\begin{figure}    
\centerline{\hspace*{0.16\textwidth}
               \includegraphics[width=1.1\textwidth,height=0.6\textheight,clip=]{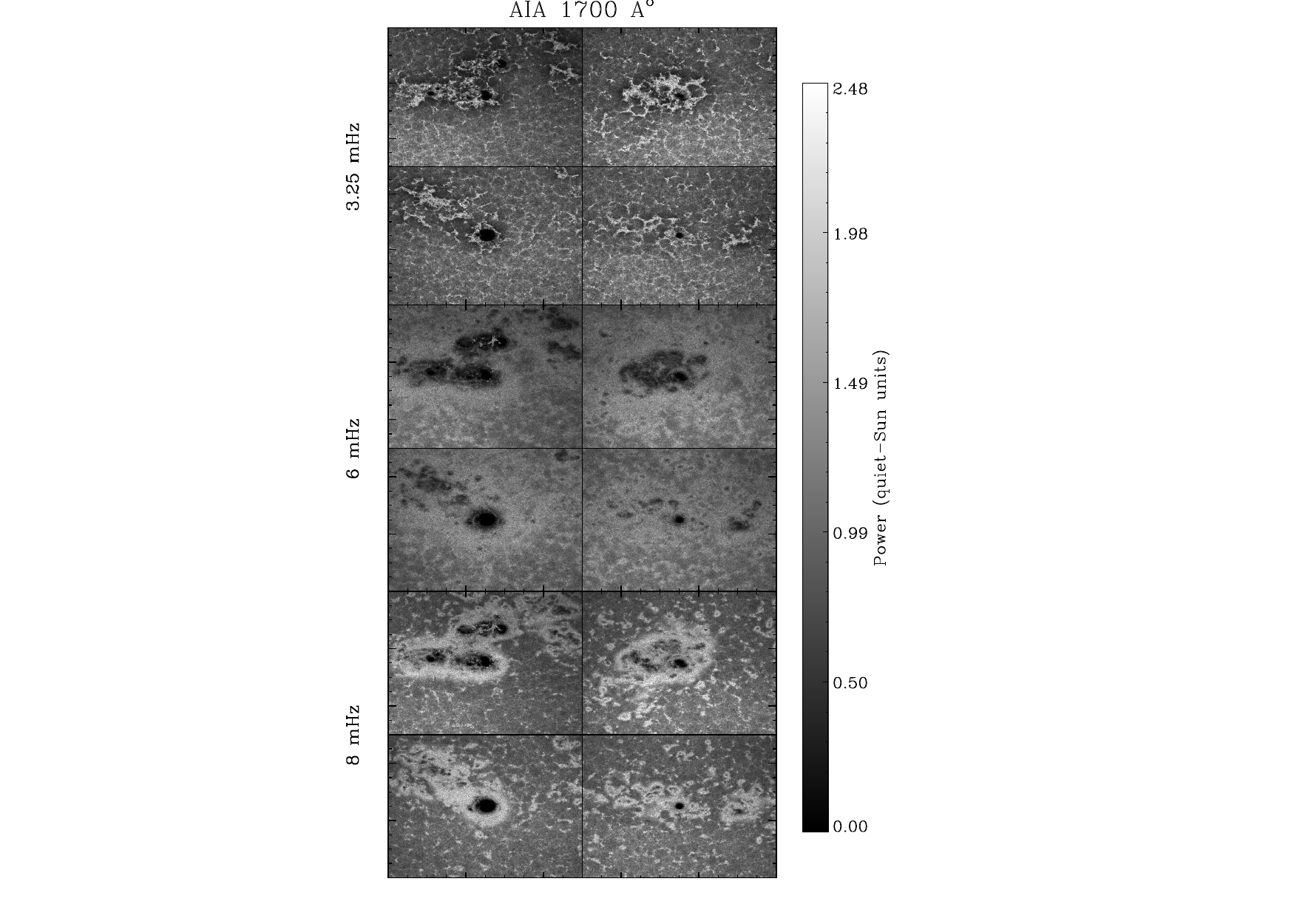}
               \hspace*{-0.65\textwidth}
               \includegraphics[width=1.1\textwidth,height=0.6\textheight,clip=]{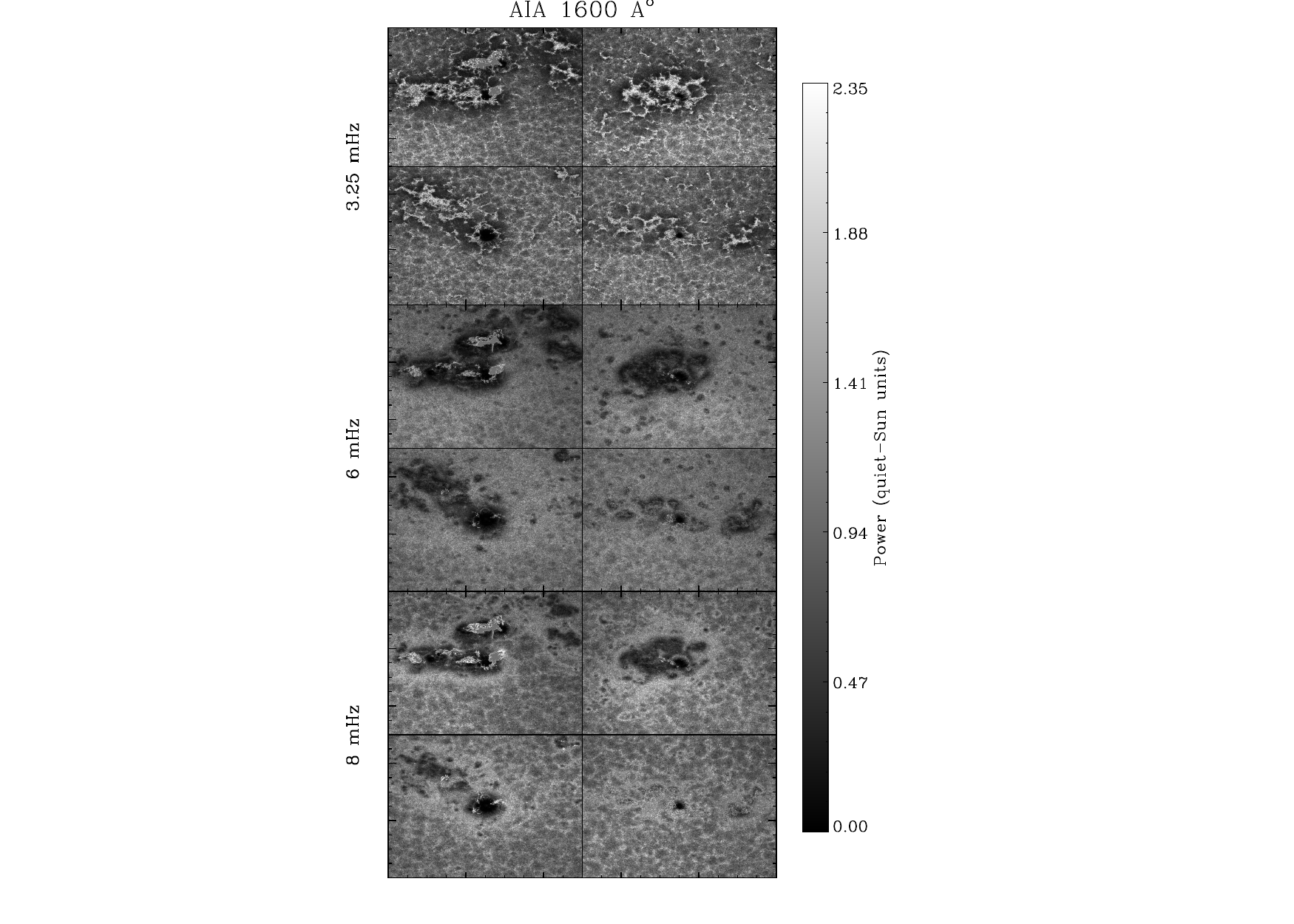}
              }
     \vspace{-0.986\textwidth}   
     \centerline{\Large \bf     
      \hspace{0.15 \textwidth}  \color{black}{(a)}
      \hspace{0.375\textwidth}  \color{black}{(b)}
         \hfill}
     \vspace{0.93\textwidth}    
\caption{Normalized power over the four active regions (arranged as in Figure 1) at three
representative frequencies, 3.25, 6, and 8 mHz (top, middle, and bottom panels, respectively) obtained from observables
AIA 1700 \AA~[$I_{\rm{uv1}}$, (a)] and 1600 \AA~[$I_{\rm{uv2}}$, (b)]. Each active region covers a square area of 373 $\times$ 373 Mm$^{2}$.} 
\label{fig4}
\end{figure}

\section{Magnetic Field and High-frequency Power Enhancements}
\label{sec.main}
Almost all previous observational studies of high-frequency power enhancements around active regions have analysed them as a 
function of LOS magnetic field, except for Schunker and Braun (2011), who used magnetic-field inclinations derived from
potential-field extrapolations of MDI-LOS magnetograms. Here, we use more direct measurements of the vector field
from HMI to analyse the power enhancements around sunspots and active regions as a function of absolute total field [$B=|\bf{B}|$]
and inclination [$\gamma$]. Since we are interested in studying wave power around sunspots, not within them, we mask out areas
within sunspots: pixels falling within the outer-penumbral boundary defined by 0.92 $I_{\rm{c}}$ are excluded from the analyses. 
This eliminates regions with $B$ stronger than about 850 G, and hence we do not study the highly suppressed power seen 
within sunspots as well as other features such as three-minute umbral oscillations and penumbral {\it p}-mode power due to wave 
propagation caused by the reduced acoustic cut-off frequency in inclined fields (Rajaguru {\it et al.} 2007, 2010).
\begin{figure}    
\centerline{\hspace*{0.03\textwidth}
               \includegraphics[width=1.02\textwidth,height=0.54\textheight,clip=]{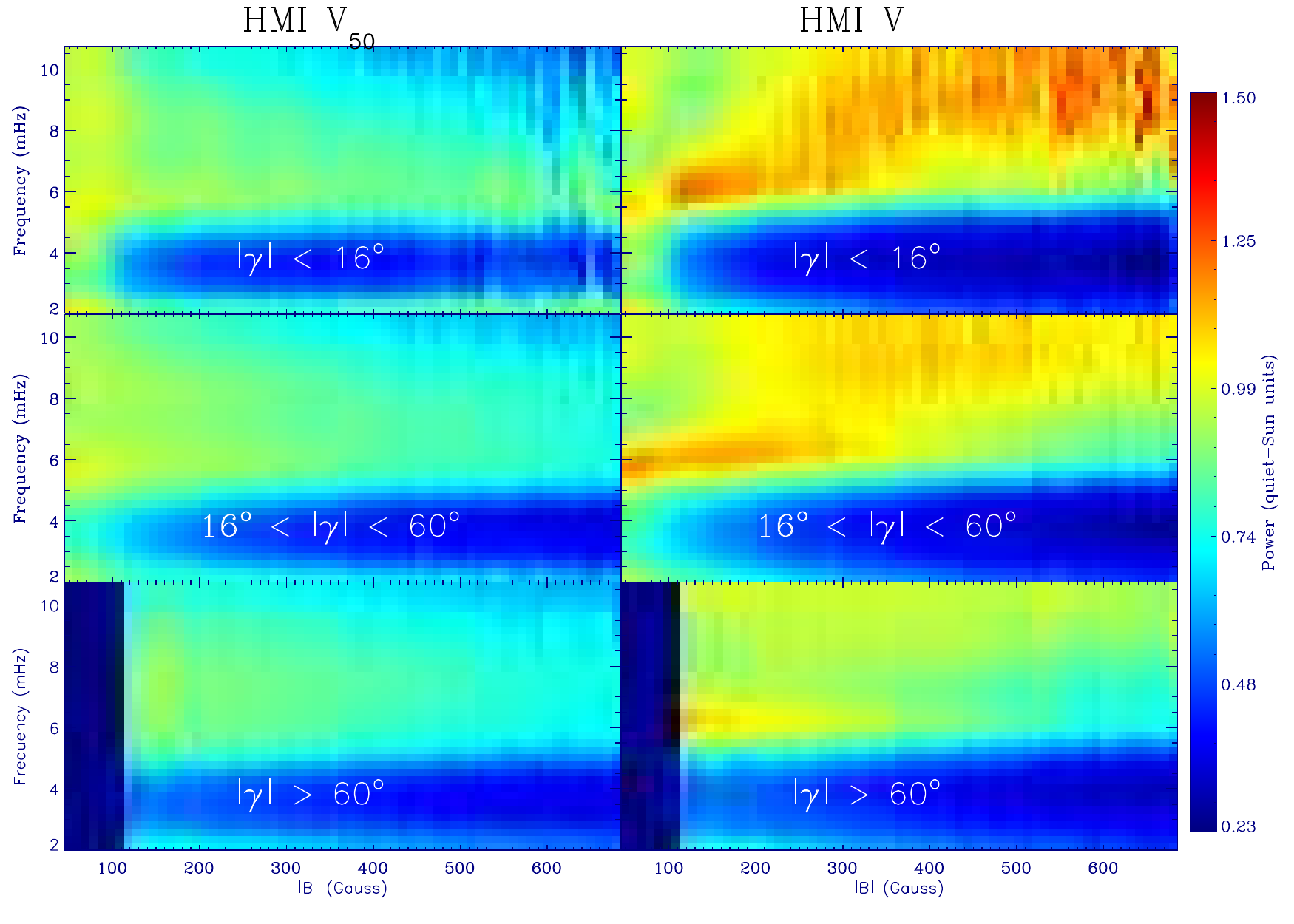} }
\caption{Power, averaged over three different ranges of $\gamma$ (as marked in the panels), as a function of total magnetic-field 
strength [$B$], for all the four active regions combined. Left column is from observable HMI $v_{50}$ and the right
from HMI $v$. Each active region covers a square area of 373 $\times$ 373 Mm$^{2}$.}
\label{fig5}
\end{figure}

Normalized power maps, {\it i.e.} power in quiet-Sun units, are averaged over 10 G bins in $B$ and 4\degree bins in $\gamma$.
For ease of analysis and appreciation of major features in the variation of halo power against $B$ and $\gamma$, we produce
two sets of figures: the first set (Figures 5, 7, 9) shows power against $B$, with pixels
grouped in three different ranges of $|\gamma|$, $viz.$ nearly horizontal field with $|\gamma| < 16\degree$, inclined fields
with $ 16\degree < |\gamma| < 60\degree$, and nearly vertical fields with $|\gamma|> 60\degree$; 
the second set shows the (Figures 6, 8, 10) dependences of power on $\gamma$, with pixels grouped in 
three different ranges of $B$, {\it viz.} $B < 100$ G, $100 < B < 200$ G, and $200 < B < 450$ G. 
We note here that $B$, in the average over the period of observation, has a noise background of about 40 G in magnitude,
and hence this value is a rough minimum for $B$.

We add a caveat here that the analysis of power halos in terms of photospheric values for $B$ or $\gamma$ do not capture 
accurately the positions of peak power in $B$ -- $\nu$ or $\gamma$ -- $\nu$ space for low values of $B$ and $|\gamma|$,
especially for observables representing the higher layers. For example, it is obvious from the spatial power
maps in Figures 2 -- 4 that enhanced power does appear over very weak- or non-magnetic pixels ({\em i.e.}, those with minimum values for
$B$ or $|\gamma|$) around sunspots or between opposite polarity regions. In this situation, the process of averaging pixels falling 
within certain small bin sizes of $B$ leads to incorrect $B$-dependence of peak power over the low end values of $B$: 
far away quiet-Sun pixels with no excess power at all get mixed with those that are non-magnetic but with enhanced power; 
this should however not affect the variations seen, say, above about $B$=50 G, or $|\gamma| >$ a few degrees.
A more accurate and physically meaningful association of wave power observed at a given height requires $B$ and $\gamma$ 
values for this height from appropriate magnetic-field extrapolation modelling. For our analysis here,
we focus on several other interesting properties of power maps that are not affected by the above finer issues, and we
defer the above sophistication to a future analysis.
\begin{figure}    
\centerline{\hspace*{0.03\textwidth}
               \includegraphics[width=1.02\textwidth,height=0.54\textheight,clip=]{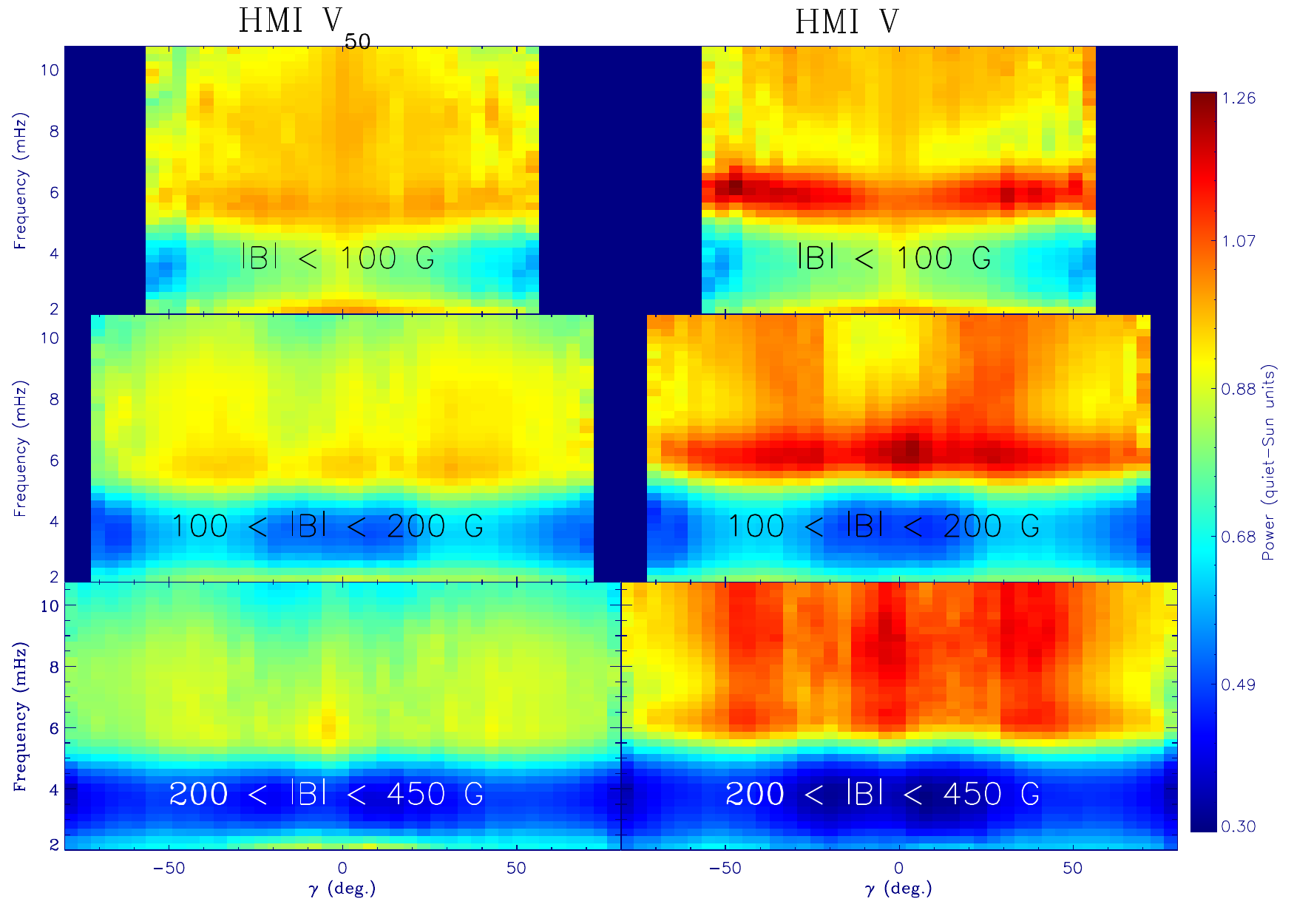} }
\caption{Power, averaged over three different ranges of $B$ (as marked in the panels), as a function of $\gamma$
for all the four active regions combined. Left column is from observable HMI $v_{50}$ and the right from HMI $v$.
Each active region covers a square area of 373 $\times$ 373 Mm$^{2}$.}
\label{fig6}
\end{figure}

\subsection{Photospheric Behaviour of Power Halos}
\label{subsec1.main}
The lowest height observables $I_{\rm{c}}$ and $v_{50}$, as seen from Figures 2a, 3a, and 5 -- 8, do not show any appreciable high-frequency
halos: while $I_{\rm{c}}$ shows no excess power at all, $v_{50}$ shows a very small excess, less than 1\%. The power maps from $v_{50}$,
especially at $\nu$ = 6 and 8 mHz, are modulated by some large- and small-scale fringes, which are of instrumental origin:
the procedure of estimating $v_{50}$ from $I_{0}$ and $I_{5}$, calibrating and tracking it is different from the standard HMI 
pipeline procedure for $v$, and these seem to cause the CCD flat-field leak into the final velocity estimates.
We have so far been unsuccessful in removing them. These fringes of instrumental origin, as can be clearly seen in Figure 2a, appear
all over and hence do not cause any confusion in the identification and analyses of halos in and around the magnetic regions.
Our main result from $v_{50}$, {\it viz.} that the power halos have very small magnitude, appears sound: since fringes modulate the
power distribution in equal and opposite magnitudes they do not contribute any net artificial signals in our analysis of real solar
signals. 

It is interesting to note that there is a broad and uniform distribution, against wave frequency [$\nu$] (Figures 
5\, --\,8) above about 5 mHz as well as against $\gamma$, of power seen in $v_{50}$ and $I_{c}$ over weak
($B <$ 100 G) field and  quiet-Sun areas. This feature indicates a more or less uniform injection of high-frequency wave power 
at the lowest photospheric heights. These high-frequency waves arrive directly from acoustic sources associated with the convective 
turbulence just below the photosphere, and provide an uniform background of wave energy above the acoustic cut-off, in 
agreement with the standard picture of propagation through the atmosphere of waves with such frequencies. 
However, as can be seen in wave-power measurements from $v$ (Figures 2b, 5, and 6), which corresponds to a height of about 140 km, 
the magnetic field profoundly modifies wave propagation between heights $z =$ 0 and 140 km and causes power excess, 
relative to non-magnetic areas at the same height level, depending on $B$ and $\gamma$: for the familiar and well-observed
halo in the $\nu$ range of 5.5 -- 7 mHz,  power over weak fields ($B \le $100 G)
is a slightly increasing function of $|\gamma|$ (see the top-right and middle-right panels in Figure 6 and 5, 
respectively) and over the intermediate-strength (100 $ < B < $ 300 G) fields it
decreases in amplitude as $|\gamma|$ increases (from top to bottom panels in the right of Figure 5), 
{\it i.e.} more horizontal the field the larger is the power excess. Overall, the maximum excess is in the latter of the above, 
and this property has been the most known one from earlier studies. In addition,
as $B$ increases, between 40 and 250 G, the excess power slowly shifts to higher frequencies, and this is the behaviour 
seen and reported by Schunker and Braun (2011). We associate the weak $B ( \le$ 100 G) inclined field ($|\gamma| >$ 16\degree)  
behaviour to the small-scale flux elements of plages and network flux tubes.
 
\begin{figure}    
\centerline{\hspace*{0.03\textwidth}
               \includegraphics[width=1.02\textwidth,height=0.54\textheight,clip=]{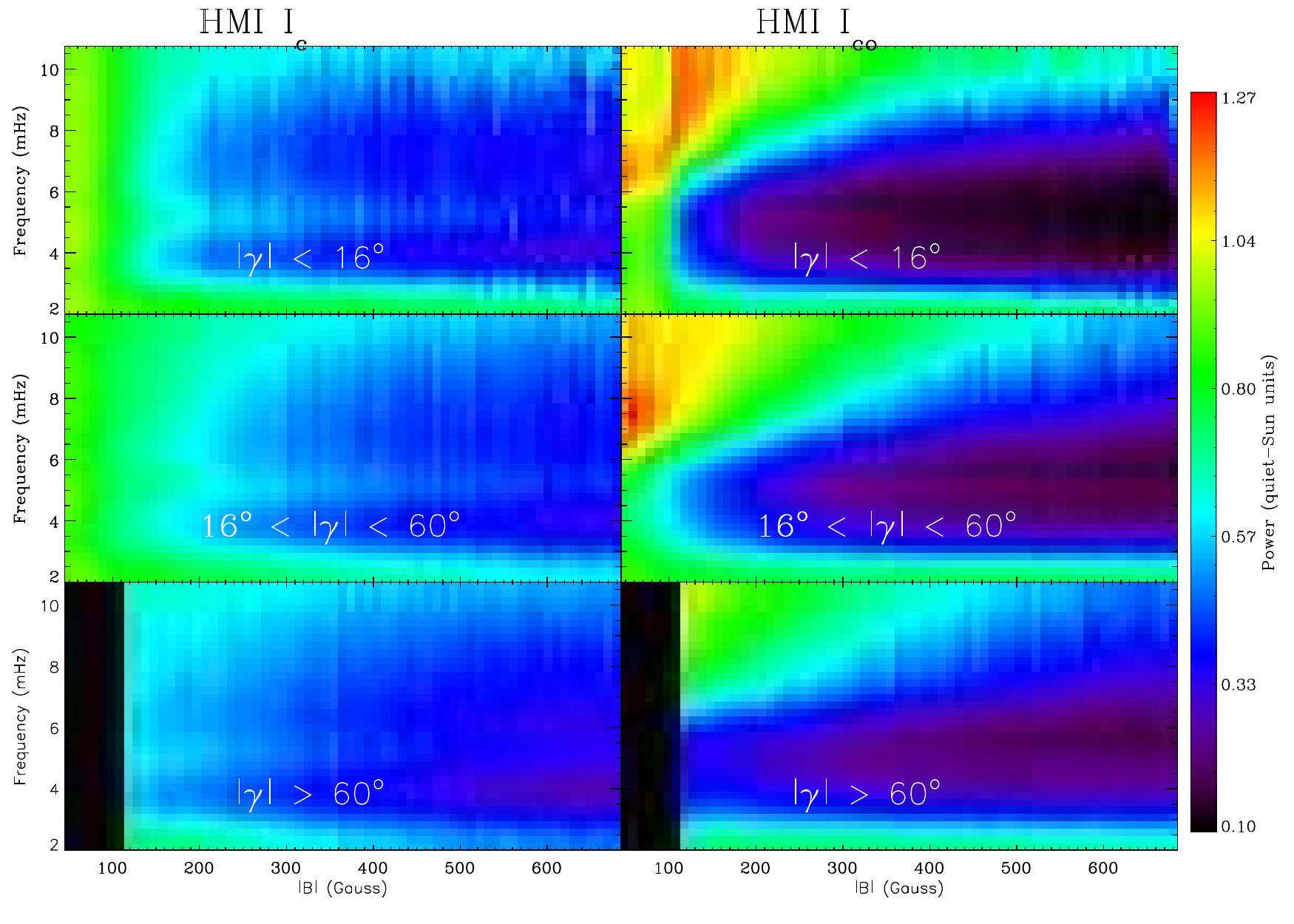} }
\caption{Power, averaged over three different ranges of $\gamma$ (as marked in the panels), as a function of total magnetic-field
strength [$B$], for all the four active regions combined, from observables HMI $I_{\rm{c}}$ (left column) and HMI $I_{\rm{co}}$ (right column).}
\label{fig7}
\end{figure}

An interesting new feature, revealed by the higher-cadence HMI velocity data 
for the first time, is a branch of the power halo in the frequency range 7 mHz and above with peak values at about 8 mHz.
This higher-$\nu$ halo connects to the 6-mHz halo between 200 and 300 G, and its frequencies depend on $B$ more sharply in the 
above $B$ range (see Figures 5 and 6, right column). Here again, the halo is prominent when $|\gamma| <$ 16\degree~ but 
at $B >$ 250 G, $i.e.$ in stronger horizontal fields at the base regions of 
canopy surrounding large structures such as sunspots; these compact halos are distinct 
in the spatial maps (Figure 2b, bottom panel). However, as seen in Figure 6, power above 7 mHz exhibits 
slightly more complex $\gamma$- and $B$-dependence in inclined field areas (20\degree $<|\gamma| < 60\degree$) when $B >$ 100 G:
the excess power shifts to less inclined fields as $B$ increases, and these are seen as two pillars of power
(corresponding to the two polarities with same $\gamma$, $i.e.$ at $\pm \gamma$) that shift to larger $|\gamma|$ as
$B$ increases. We identify this latter $|\gamma| > 20\degree$ halos as those surrounding the small-scale flux elements,
which possibly do not fan out too strongly to produce large horizontal fields at heights
where $v$ is measured. If we associate the locations of power halos, in height [$z$] and $B$, to layers where the
plasma $\beta \approx$ 1 (see Section 4), then the above features (in Figures 5 and 6) would 
indicate that these layers coincide with large horizontal-field locations surrounding sunspots while the same would
occur over similar height ({\em i.e.}, about the same strength) but inclined fields (20\degree $<|\gamma| < 60\degree$) of 
small-scale flux tubes. This, in turn, would imply slightly slower rate of expansion of small-scale flux tubes than the
sunspot fields. Thus, it is clear that the existence of power halos at $\nu >$ 7 mHz is dictated largely 
by $\beta$ = 1 transitions ({\it i.e.} by $B$) through the atmosphere, rather than by $\gamma$.
\begin{figure}    
\centerline{\hspace*{0.03\textwidth}
               \includegraphics[width=1.02\textwidth,height=0.54\textheight,clip=]{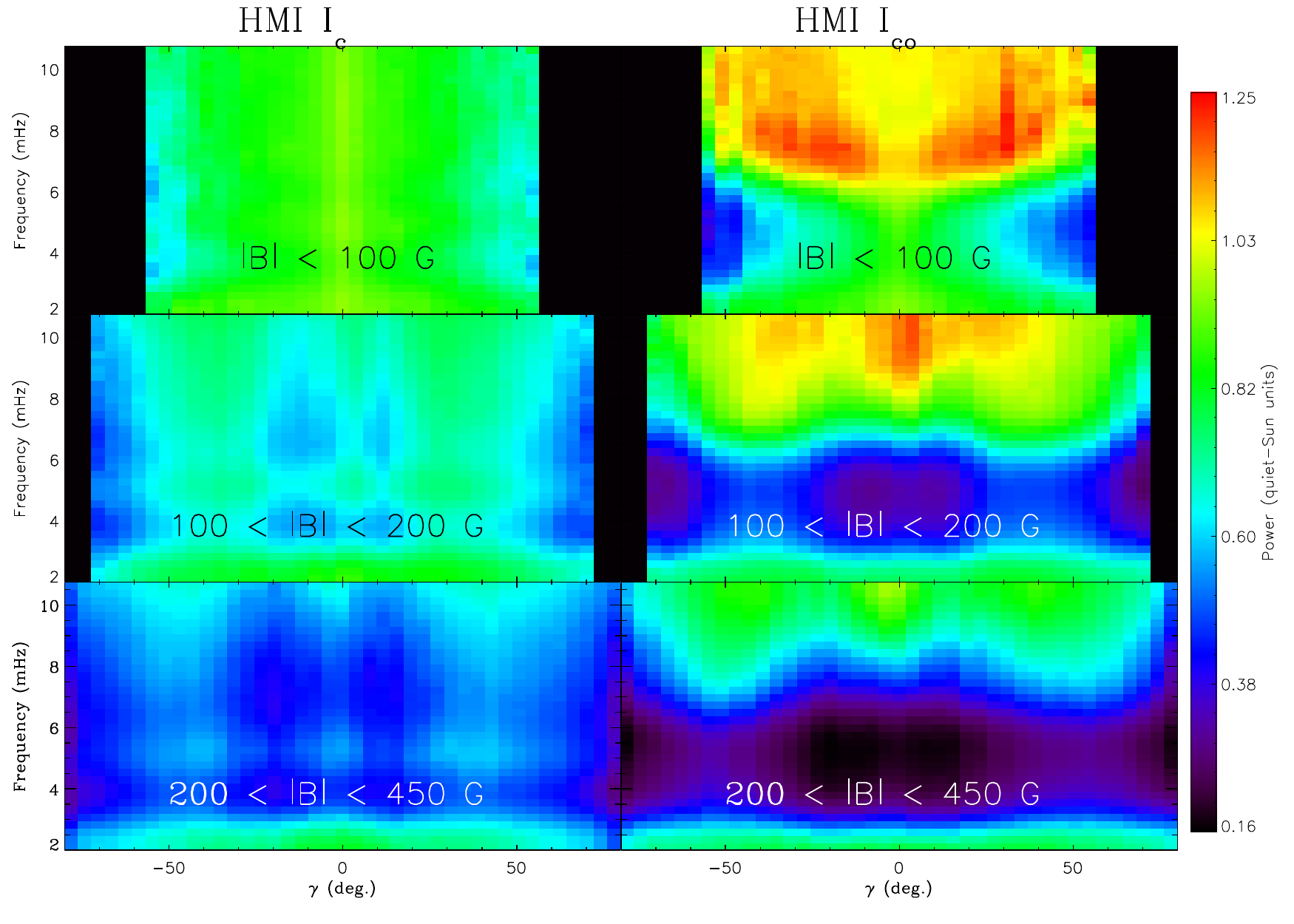} }
\caption{Power, averaged over three different ranges of $B$ (as marked in the panels), as a function of $\gamma$
for all the four active regions combined. Left column is from observable HMI $I_{\rm{c}}$ and the right from HMI $I_{\rm{co}}$.}
\label{fig8}
\end{figure}

Finally, as an interesting and important feature that possibly relates to wave refractions and associated horizontal propagation
around the $\beta$ = 1 layer \cite{khomenkoandcollados09}, we identify the reduced-power region between the outer,
much extended, weak enhancement (over $B <$ 100 G region, top-right panel of Figure 6) and the strong compact halo 
(closer to fields of $B$ = 200 G and above) discussed above.
This curious reduced-power region in spatial maps (Figure 2b, bottom panel) corresponds to, in the $B$ $vs$ $\nu$ maps of Figure 5 (top two right-side panels),
a region that starts at ($\nu, B$) $\approx$ (7 mHz, 40 G) and extends across a curved region over $B$ in the range of 100 - 200 G 
and higher $\nu$. The values of power in this region, as light green shade in Figure 5, are less than those obtained 
from $v_{50}$ over the same $\nu$ -- $B$ region (seen as yellowish region over the same location in the left panel of Figure 5). 
We reason that these features could be signatures of a refracting 
fast-magnetoacoustic wave travelling from high $\beta$ layers below \cite{khomenkoandcollados09}, 
corresponding to one or both of the following two scenerios. 
i) the intermediate $B$ = 100 -- 200 G field corresponds to $\beta$ = 1 height well above 140 km from where $v$ signals
arise, and hence no excess power at this latter height, whereas this height difference is close to zero over the stronger-field 
compact halo region closer to the spots and strong field structures; the outer much-extended weak 8-mHz halo over the weak 
field region ($B<$ 100 G) could be due to the fast- to slow-mode conversion resulting in propagation along the field lines 
that curve back towards the photosphere. ii) If the Doppler $v$ indeed capture waves from around the $B$ = 100 -- 200 G, $\beta$ = 1
layer [note that the estimate of 140 km height refers only to the peak position of response function 
for $v$ for the HMI line \cite{flecketal11}, contributions to which come from layers as high as 300 km \cite{nortonetal06}],
then the reduction in power could arise directly from the small or vanishing vertical (which is close to the LOS for the
regions studied here) velocities due to horizontal propagation in the refraction region of the fast wave. 
Increasing fast-mode speed in height leads to eventual reflecton toward the lower-
height stronger field base of the canopy causing the compact halos immediatly surrounding
the spot (bottom panel of Figure 2b). If the waves are mainly compressive, then intensity observations should not exhibit such
reductions. What we show and discuss below using power maps from $I_{\rm{co}}$, which forms closest in height to $v$, 
and from $I_{\rm{uv1}}$ and $I_{\rm{uv2}}$ (see Section 3.2) appear to support the scenerio ii).

The above-observed features may agree qualitatively with the theory 
and simulations advanced by \inlinecite{khomenkoandcollados09} that show magneto-acoustic wave refractions as possible causes of 
power enhancements, and also agree well with such wave refractions studied by \inlinecite{nuttoetal11} using 2D MHD simulations.
We discuss further these aspects and explanations provided by the theory and
simulations referred to above, based on our observed features and results, in Section 4.
 
Power maps derived from $I_{\rm{c}}$ and $I_{\rm{co}}$ are shown in Figures 3b, 7, and 8. 
In conformation with earlier results, $I_{\rm{c}}$ does not show any high-frequency power excess.
In $I_{\rm{co}}$ maps, the twin-halo structure, in $\nu$, as a whole shifts towards higher $\nu$ (clearer in Figure 8): 
the halo at lower values of $B$ and $\nu$ now centers about 7 mHz (in the range 6 -- 8 mHz), while that at higher $\nu$ and $B$ 
peaks beyond 8 mHz. Excess power at 6 mHz is now small, typically less than 10\% (middle panel of Figure 3b). 
However, both the halos taper down sharply as $B$ increases, with the 6 - 8 mHz 
halo power decreasing faster than that above 8 mHz, and no excess power is seen beyond about 250 G. It is worth noting that,
in contrast to that from $v$, the 6 -- 8 mHz halo peaks at inclined (16\degree $< |\gamma| <$ 50\degree) 
lower $B$ ($<$ 100 G) region; in addition, the excess power migrates to higher frequencies as $\gamma$ 
(in the above range) increases for weak fields (the wine-glass like structure in the top right panel of Figure 8). Interestingly,
at those $\nu$ -- $B$ locations where a reduction in power is observed in maps from $v$, {\it i.e.} between 100 and 200 G and above
about 8 mHz (in Figure 5), we now see bright
halos as measured from $I_{\rm{co}}$ (Figure 3b, bottom right panel, and Figure 7). Although it forms around 150 km above
the mean height for $v$, it is likely that these high-$\nu$ waves seen in $I_{\rm{co}}$, as well as those seen in $I_{\rm{uv1}}$ and 
$I_{\rm{uv2}}$ from AIA discussed below, are the same ones seen in $v$. This gives further credance to our earlier inference on 
wave refraction and attendant horizontal propagation made from $v$ power maps.
\begin{figure}    
\centerline{\hspace*{0.03\textwidth}
               \includegraphics[width=1.02\textwidth,height=0.54\textheight,clip=]{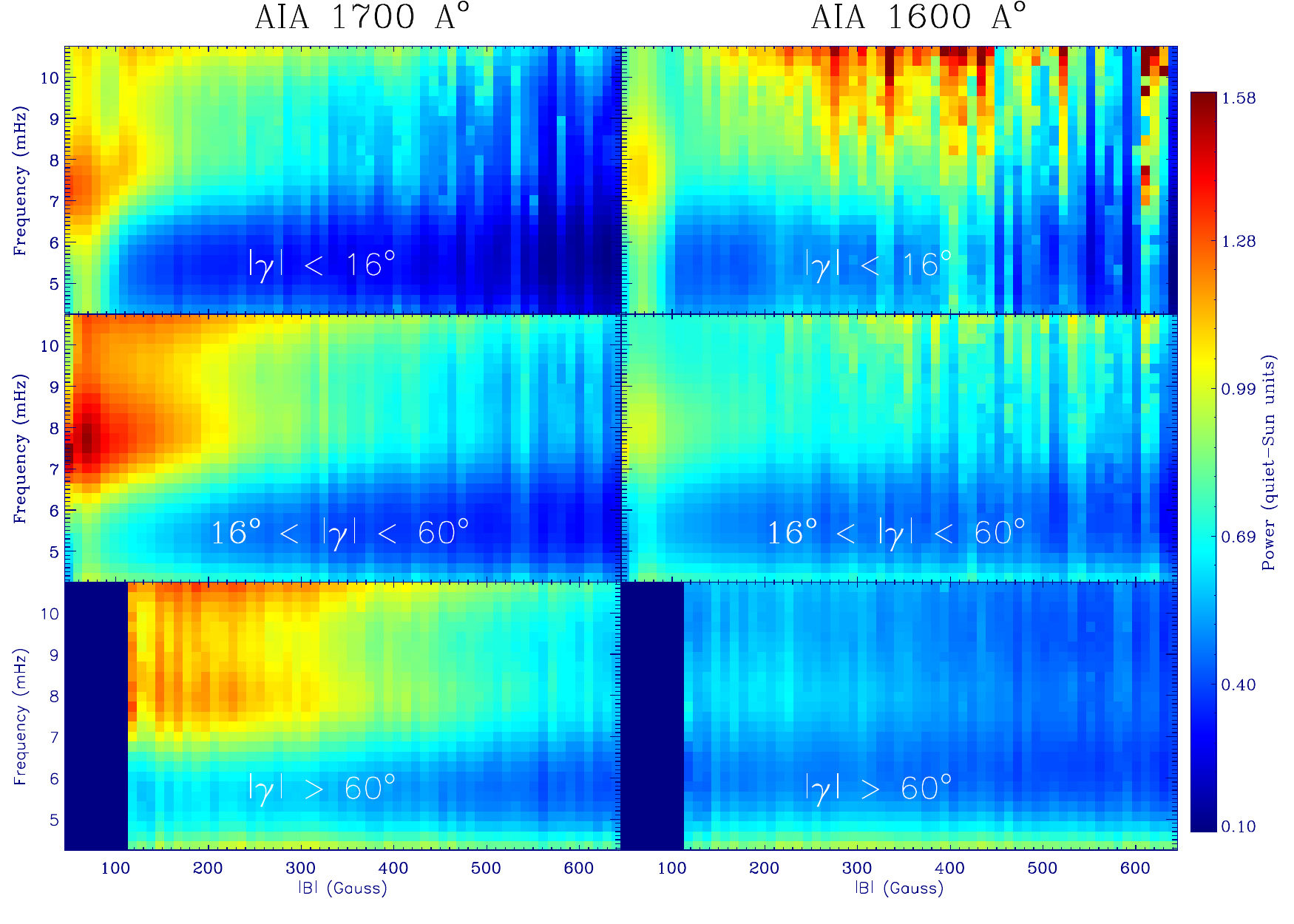} }
\caption{Power, averaged over three different ranges of $\gamma$ (as marked in the panels), as a function of total magnetic-field
strength [$B$], for all the four active regions combined. Left column is from the observable $I_{\rm{uv1}}$ and the right from $I_{\rm{uv2}}$.}
\label{fig9}
\end{figure}

\subsection{Upper-Photospheric and Lower-Chromospheric Behaviour of Power Halos}
\label{subsec:chromo}
\begin{figure}    
\centerline{\hspace*{0.03\textwidth}
               \includegraphics[width=1.02\textwidth,height=0.54\textheight,clip=]{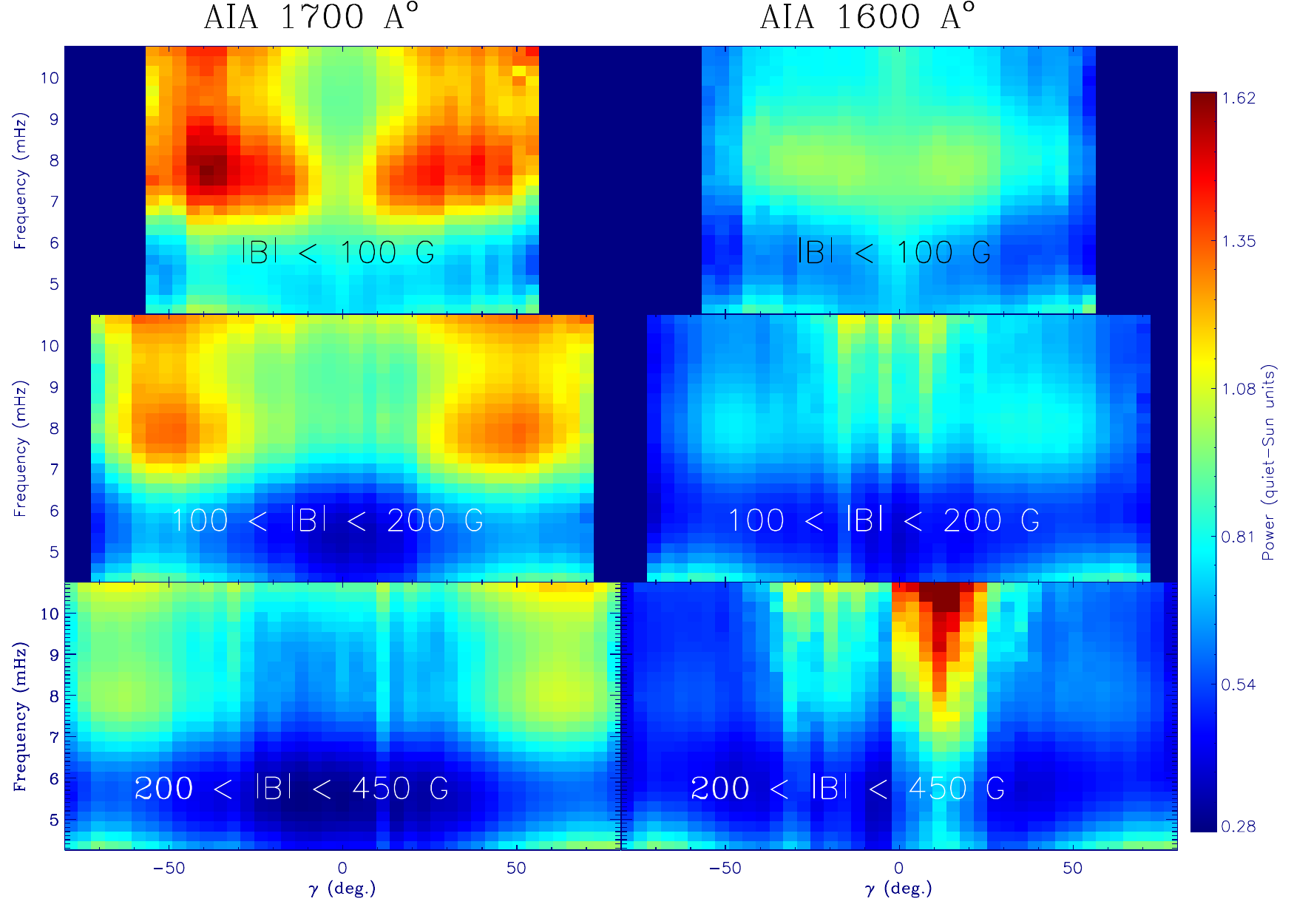} }
\caption{Power, averaged over three different ranges of $B$ (as marked in the panels), as a function of $\gamma$
for all the four active regions combined. Left column is from observable $I_{\rm{uv1}}$ and the right from $I_{\rm{uv2}}$.}
\label{fig10}
\end{figure}

Power maps estimated from the upper-photospheric and lower-chrom-ospheric UV emissions, $I_{\rm{uv1}}$ and $I_{\rm{uv2}}$, 
are shown in Figures 4, 9, and 10.
Note that the network region and plage flux tubes show prominent 5 minute oscillations (as seen in the top two panels
of Figure 4), and these oscillations extend further down to about 1.75 mHz in $\nu$, forming the so-called long period
network oscillations, known from the early 1990's \cite{litesetal93}. We do not discuss or analyse their properties here. 
Similar to that seen from $I_{\rm{co}}$, high-frequency 
power maps from $I_{\rm{uv1}}$ have the whole $\nu$-pattern of twin halos shifting upwards in $\nu$, 
with the lower end at about 6 mHz exhibiting 
a weak, but much outward extended, halos surrounding sunspots and strong flux elements. The $\nu$-extent of the
weaker-$B$ halo (see Figures 9 and 10) is now broader than that seen in $I_{co}$, covering the range of 6 -- 9 mHz, with 
the peak excess power about 8 mHz. However, $I_{\rm{uv2}}$ does not show the secondary $\nu >$ 9 mHz halo seen both in $I_{\rm{co}}$ and
$I_{\rm{uv1}}$; the apparently noisy and large-amplitude excess seen over high-$B$ (300 G and above) in Figures 9 and 10 
is due to flare-activity-induced, especially in NOAA 11161, enhancements seen to fall over the sunspots and nearby strong-
field pixels (as seen in Figure 4b, middle and bottom panels). The overall properties of 6 -- 9 mHz halo seen in $I_{\rm{uv1}}$ and
$I_{\rm{uv2}}$ are in agreement with several earlier
results on acoustic halos in the chromosphere, using the Ca II K emissions \cite{braunetal92,tonerlabonte93,ladenkovetal02}
and Na D$_{2}$ line (5890 \AA) and K D$_{1}$ line (7700 \AA) observations \cite{morettietal07}. The 1700 \AA~ emissions, $I_{\rm{uv1}}$,
exhibit the most prominent and largest amplitude excess in the 6 -- 9 mHz range. 

The close proximity, in height, of formation of $I_{\rm{co}}$ and $I_{\rm{uv1}}$ is reflected clearly in the very similar 
$B$--$\nu$ and $\gamma$--$\nu$ variations seen in the right panels of Figures 7 and 8 and the left panels of Figures 9 and 10.
In agreement with our earlier inference based on results from $v$, power over inclined fields (20\degree 
$<|\gamma| < 60\degree$) measured from $I_{\rm{uv1}}$ (Figure 10, left panel) corresponds to small-scale strong flux 
concentrations outside of sunspots, and the spatial maps (Figure 4) make this enhanced high-$\nu$ power over
such regions very obvious. 
Overall, it is clear that spatial extent of power halos and their frequency extent increase as a function of height 
in the atmosphere. This is consistent with that expected from expanding field lines and hence
upwardly raising locations of $\beta$ = 1 layer around sunspots and other flux structures.
The slow migration towards higher $\nu$ of the power halos in height in the atmosphere
may relate to the height variation of acoustic cut-off frequency and preferential conversions of waves of such frequency 
around the $\beta$ = 1 layer.
  
\subsection{Comparisons of Frequency Variation of Power over Height}
\label{subsect:compare}
A comparative picture of the variation of power against $\nu$, as well as observation height, is better seen by plotting vertical
cross-sections of Figures 5 -- 10 averaged over representative ranges of $B$ and $\gamma$. This is done in Figure 11,
for $|\gamma| <$ 16\degree (panel a) and 16\degree$ < |\gamma| <$ 60\degree (panel b). The error bars plotted are standard
errors estimated assuming that each measurement over all pixels falling within the bin sizes used for $B$ and $\gamma$ captures
an independent realisation of the same wave process determined by the values taken by the above physical quantities over the
set of pixels and hence that the scatter within such bins are random.
For clarity, we have plotted error bars only at every 1 mHz.
\begin{figure}    
\centerline{\hspace*{0.06\textwidth}
               \includegraphics[width=1.00\textwidth,height=0.450\textheight,clip=]{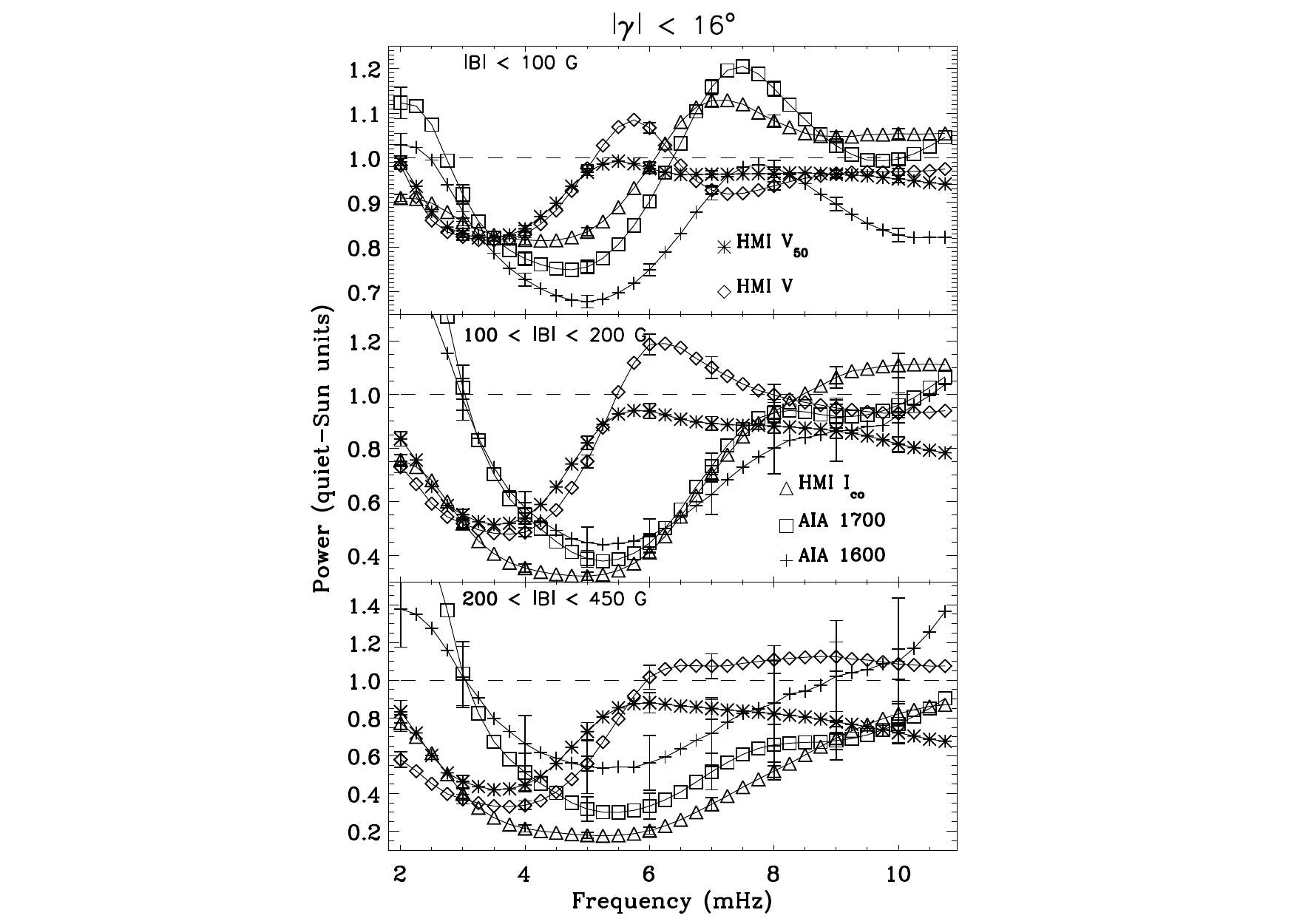}
               \hspace*{-0.5\textwidth}
               \includegraphics[width=1.00\textwidth,height=0.450\textheight,clip=]{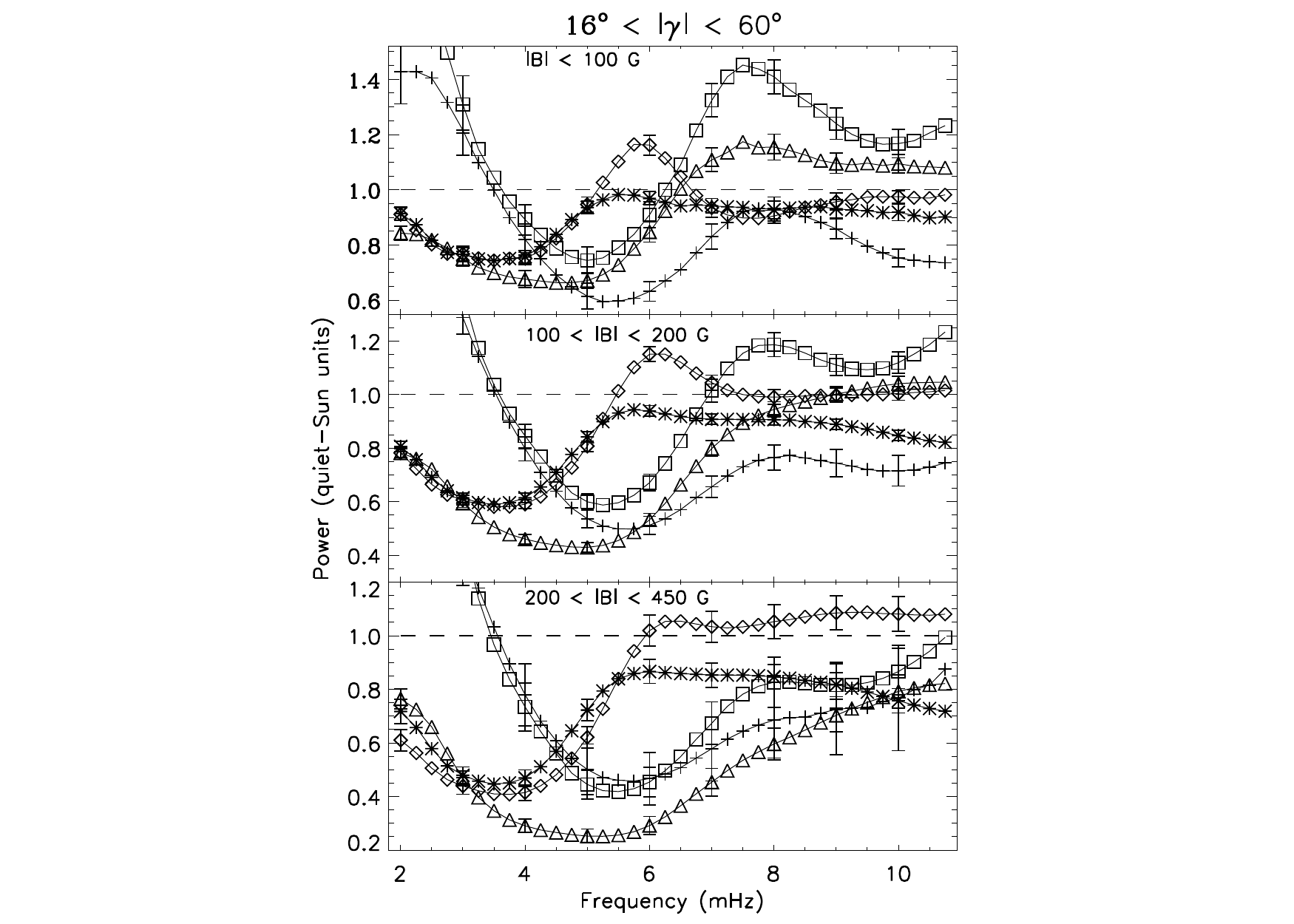}
              }
     \vspace{-0.74\textwidth}   
     \centerline{\Large \bf     
      \hspace{0.06 \textwidth}  \color{black}{(a)}
      \hspace{0.355\textwidth}  \color{black}{(b)}
         \hfill}
     \vspace{0.70\textwidth}    
\caption{Cuts against $\nu$, of Figures 6, 8, and 10 averaged over $\gamma <$ 16\degree (panel $a$) and
16\degree $< \gamma <$ 60\degree (panel $b$). The different symbols correspond to different observables used in this study,
and are marked in panel ($a$)}
\label{fig11}
\end{figure}

The more or less uniform values of power, with values close to one without any halo, above the acoustic cut-off of 5.3 mHz,
in the measurements from $v_{50}$ corresponding to the lowest heights (plotted as connected asterisks in Figure 11),
especially over weak field ($B <$ 100 G) regions is clear. The migration of peak power, for the observable $v$ (plotted
as connected diamonds), to higher $\nu$ as $B$ increases is clear. Our new finding of the
existence of a spatially compact halo in maps from $v$, at $\nu$ above about 6.5 mHz with quite a broad peak, is seen in the
bottom panels (for 200 $< B <$ 450 G) of Figure 11.

Overall, the largest amplitude power excess is observed from $I_{\rm{uv1}}$ over inclined-field 
regions (16\degree $< \gamma <$ 60\degree ) peaking at $\nu \approx$ 7.5 mHz, and this is due to the fact that, at these heights, 
all magnetic structures, small and large, are surrounded by bright halos. It is also to be noted that power-halo amplitudes,
as well as their $\nu$ variation, are very similar for $I_{\rm{co}}$ and $I_{\rm{uv1}}$ (squares and triangles in Figure 11, respectively), 
reinforcing our overall inference that variation of magnetic field (or $\beta$ = 1 layer) over height in the atmosphere 
controls the wave physics behind the power excess. 

\subsection{Wavenumber Dependence of Power Spectra over Height}
\begin{figure}    
\centerline{\hspace*{0.08\textwidth}
               \includegraphics[width=0.87\textwidth,height=0.450\textheight,clip=]{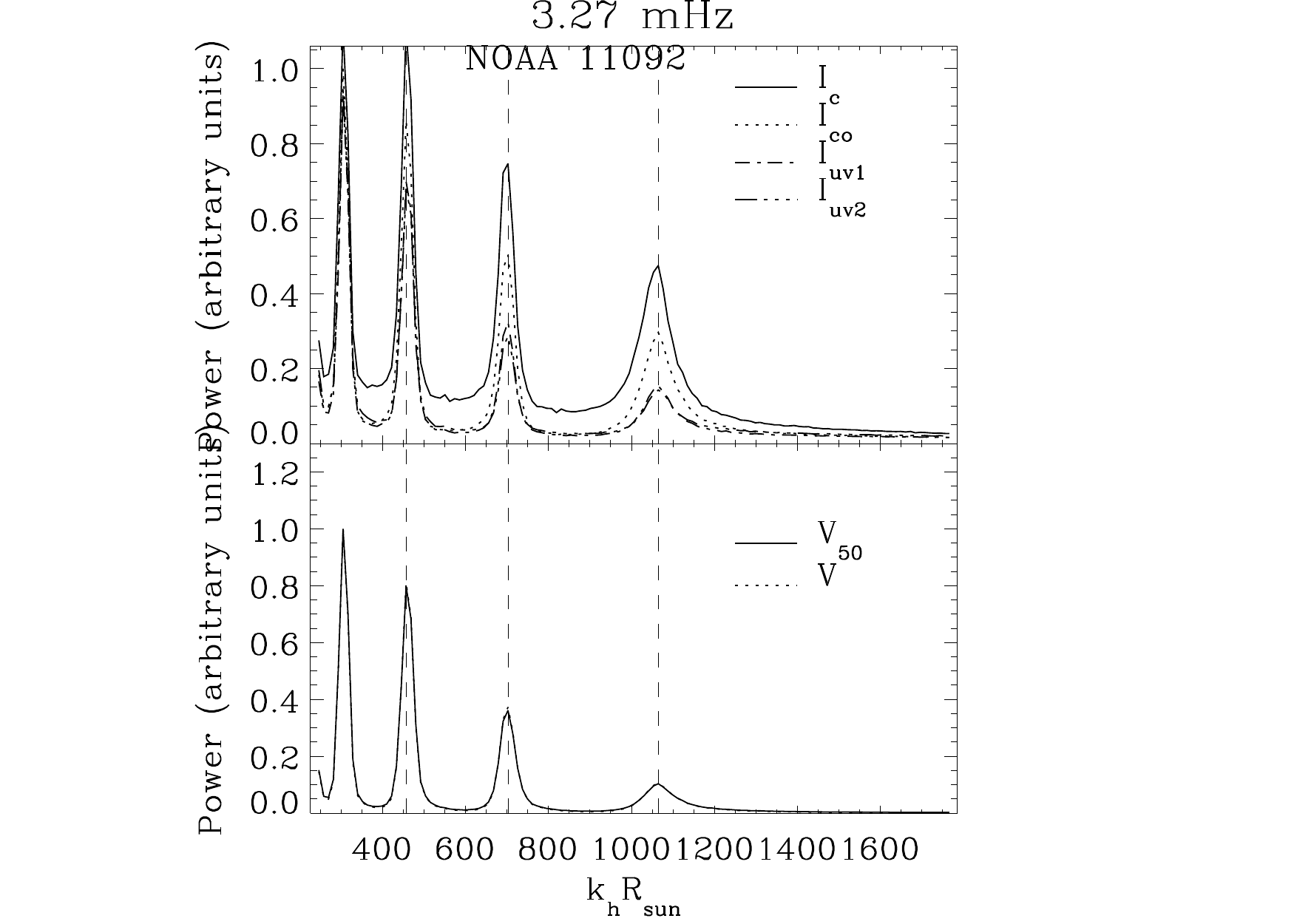}
               \hspace*{-0.36\textwidth}
               \includegraphics[width=0.87\textwidth,height=0.450\textheight,clip=]{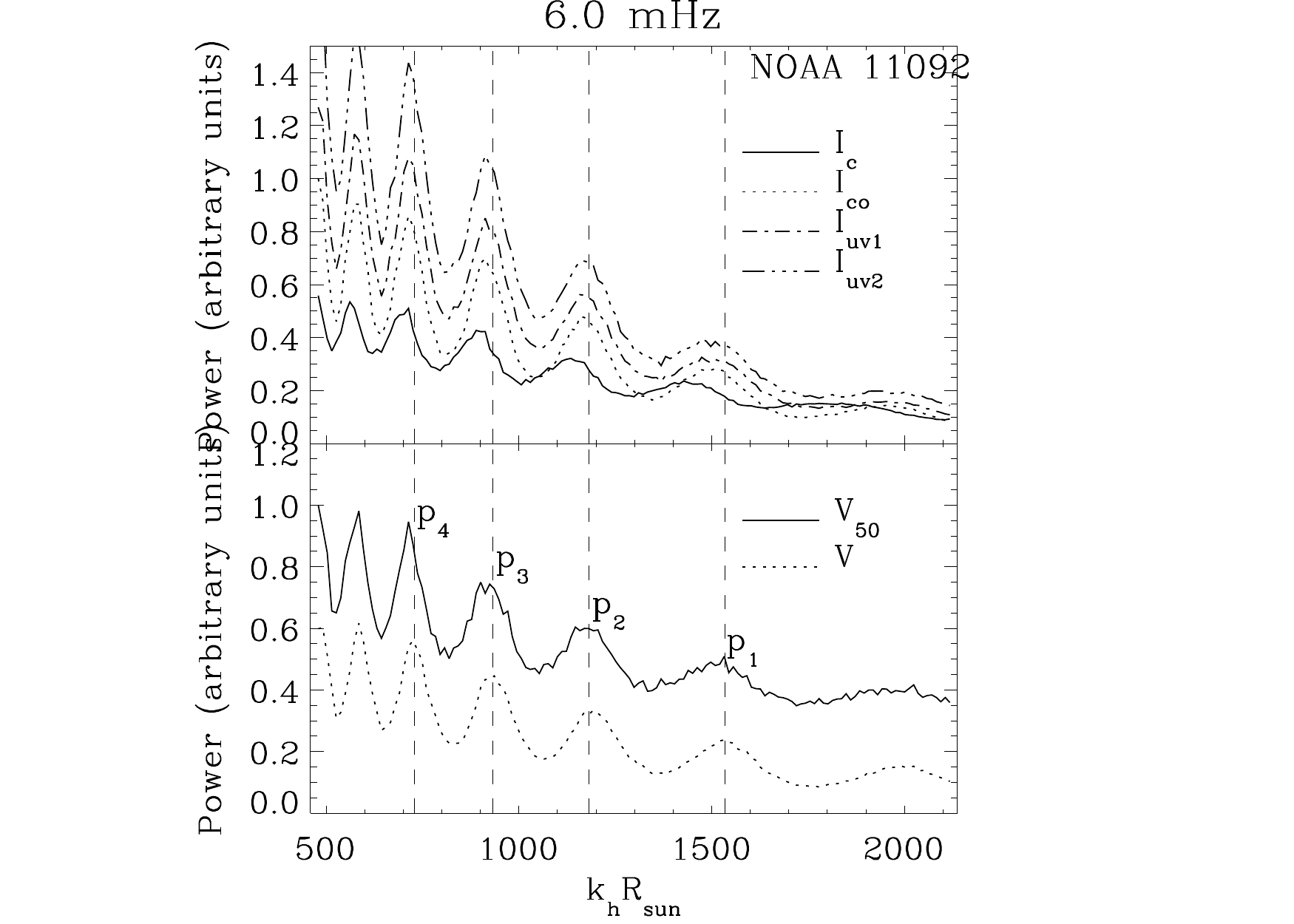}
              }
     \vspace{-0.74\textwidth}   
     \centerline{\Large \bf     
      \hspace{0.06 \textwidth}  \color{black}{(a)}
      \hspace{0.355\textwidth}  \color{black}{(b)}
         \hfill}
     \vspace{0.70\textwidth}    
\caption{Modal power spectra against $k_{\rm{h}}$ at two different values of $\nu$ for the different observables used in this
study (different line styles correspond to different observables as marked within each panel): 
(a) for $\nu$ = 3.27 mHz, and (b) for $\nu$ = 6 mHz. Top row panels correspond to spectra from intensities and the bottom
panels to those from velocities. See text for further explanations.}
\label{fig12}
\end{figure}
It is well known that the solar oscillation power spectra exhibit ridge structure well beyond the photospheric
cut-off frequency of $\approx$ 5.3 mHz \cite{libbrecht88}, and that it is not due to the resonant p modes 
as in the interior but are understood to be due to an atmospheric interference between waves directly from the 
sub-surface sources and those refracting back from the solar interior \cite{kumaretal90,kumarlu91}.
The high-frequency power halos
are reported to introduce slight shifts in the locations of ridges in the power spectra \cite{schunkerandbraun11}:
at a given frequency, active regions shift the wavenumbers of high-frequency waves to slightly higher values
as compared to quiet Sun; higher wavenumber for a given frequency means that the peak-power locations against frequency
({\it i.e.} the ridges) shift downwards and hence reduced frequencies. Here, we check if changes occur for the same active region 
as a function of observation height ({\it i.e.}, as a function of different observables used here).

We azimuthally average the three-dimensional ($k_{x}, k_{y}, \nu$) power spectra at each $\nu$ to make $k_{\rm{h}}$-$\nu$ spectra.
We show the results for one active region, NOAA 11092, in Figure 12. We produce the $k_{\rm{h}}$ variation of the spectra
at two frequencies: 3.27 and 6 mHz. The spectra estimated from all intensities, $I_{\rm{c}}$, $I_{\rm{co}}$, $I_{\rm{uv1}}$, and $I_{\rm{uv2}}$
are grouped together (top panels), and those from the velocities $v_{50}$ and $v$ are shown in the bottom
panels of Figure 12. Vertical-dashed lines across both set of panels mark locations of lowest radial order p modes as determined by
the peak locations of power in $v$ (dotted lines in the bottom panels). It is clear that 
at $\nu$ = 3.27 mHz, there is no change in the wavenumber dependence as a function of height (Figure 12a) and this
is expected from the overall evanescent nature of such frequency waves. At $\nu$ = 6 mHz (Figure 12b), the major change is a 
shift towards lower $k_{\rm{h}}$ of the locations of power peaks obtained from $I_{\rm{c}}$ than the rest of the observables. This
change in power spectra corresponds to the well observed positive shifts, {\it i.e.} increase in frequencies,
related to the ``correlated noise" mechanism that explains the reversal of asymmetry in $p$-mode power spectra from intensities
in relation to that from velocities \cite{nigametal98,kumarbasu99}. These increased frequencies observed
in $I_{\rm{c}}$ correspond to the decrease in $k_{\rm{h}}$ seen in Figure 12b, and hence could not be due to the absence of power halos
in $I_{\rm{c}}$. However, other intensities $I_{\rm{co}}$, $I_{\rm{uv1}}$ and $I_{\rm{uv2}}$, which are from progressively increasing heights
in the atmosphere, do not show these shifts and have power peaks at almost the same positions as those from $v$. This again
possibly relates to the decreasing correlated noise as distances from the acoustic sources increase for these intensities
from higher layers. Hence, the above changes in power-peak locations are possibly not due to the power halos {\it per se}.
Additionally, comparing $v_{50}$ and $v$ power spectra in Figure 12b (bottom panel), the increasing high-frequency power halo 
in $v$ does not appear to leave any significant changes in the power spectra, although there are slight shifts noticeable
in the locations of $p_{3}$ and $p_{4}$ ridges. However, such slight changes, with respect to $v$, are also seen in higher
atmospheric intensities (top panel of Figure 12b). Thus, we can conclude that there are no clear associations of variations 
in high-frequency power excess against height with any significant changes in the power-peak locations.
The detected changes, in comparison with non-magnetic regions, as reported by \inlinecite{schunkerandbraun11} are of larger
magnitude than any changes caused by variations in power halo against height.

A thorough analysis of differences in the spectral power-peak locations and asymmetries in power profiles, from among those
obtained from intensities at different heights, and their relationship to the physical mechanisms controlled by correlated
noise and acoustic source depths \cite{nigametal98,kumarbasu99} is beyond the scope of the work reported here, and we focus on
this in a separate study to be reported in the near future. 

\section{Discussion and Summary}
\label{sec:discuss}

The physics of interactions between acoustic waves and magnetic fields in the solar atmosphere leads to a rich variety 
of observable dynamical phenomena, the understanding of which is crucial to mapping the thermal and magnetic structuring
in height of the solar atmosphere. A large body of theoretical and observational studies of such physics exists
and a significant fraction of which, as explained in Section 1, is relevant for the detailed analyses
we have made here of the high-frequency power halos around active regions. The spatial reorganization of otherwise
uniform injection of acoustic waves of frequency above the photospheric cut-off of $\approx$ 5.3 mHz from the subsurface
and photospheric layers by the overlying and arching magnetic field of active regions and strong field flux elements
is clearly brought out in our analyses of wave power distribution estimated from observables spanning a height range
of 0 -- 430 km above the photosphere. In accordance with the central theme of theoretical investigations referred to in 
Section 1 ({\it e.g.} \inlinecite{rosenthaletal02,bogdanetal03,khomenkoandcollados09}, see also the review by
\inlinecite{khomenko09}), our results presented and discussed in detail in Section 3 clearly bring
out the importance of interactions between fast-acoustic waves from the lower atmospheric high-$\beta$ regions 
and the expanding and spreading magnetic field canopy that separates the low-$\beta$ region above. 
Although we have not modeled the higher atmospheric magnetic field from the observed vector field at photospheric heights
and have not estimated and mapped the $\beta$ = 1 layer, results obtained from an analysis of photospheric $B$- and $\gamma$-dependence 
of power maps, especially for the newly identified high-frequency secondary power-halo peaking at about
$\nu$ = 8 mHz (Figures 2(b), 5, and 6), appear to agree with the theory and simulations performed by 
\inlinecite{khomenkoandcollados09}: refraction and subsequent reflection of fast magneto-acoustic waves around the 
$\beta$ = 1 layer as agents of additional wave energy deposition at these layers and hence the power excess.
Such refracted fast magneto-acoustic waves, possibly depending on the angles of incidence of acoustic waves prior to 
conversion, may converge towards the stronger-field base region that fans out from the sunspots and other 
strong-field structures, and cause a focussing effect thereby increasing wave amplitudes and thus the compact halos 
seen in maps from $v$ (Figure 2(b)). The reduction in power encircling the compact halos itself is again identified as a signature
wave refraction under two possible scenarios: i) the refraction and horizontal propagation region ({\it i.e.}, 
the intermediate 100 -- 200 G field $\beta$ = 1 layer) being located at a height significantly above the $v$-formation height (of 140 km)
or/and ii) vanishing velocity signals in the vertical direction due to horizontal propagation around the observation height (for $v$).
The spatially extended and weak halo surrounding this reduced power region
is suggested to result from the field-aligned slow magneto-acoustic wave due to the fast-to-slow mode conversion.
In regions where opposite polarity canopy fields meet to produce neutral lines, it is conceivable that both the refracted
fast waves and converted slow waves moving in opposite directions toward the neutral lines dump or focus wave energy causing 
greater power seen at these locations \cite{schunkerandbraun11}.

Other major results of this study can be summarised as follows: i) visibility of enhanced power is a strong function of
height in the atmosphere; absence of power excess in continuum intensity [$I_{\rm{c}}$] is due to the corresponding
height being significantly below the wave-mode conversions happening around the $\beta$ = 1 layers, and hence not due to
any incompressive wave; this is confirmed by the presence of strong halos in maps from
line core intensity $I_{\rm{co}}$, which form at higher layers. ii) The well-observed 6 mHz halo is the strongest in maps
from Doppler velocities [$v$] forming at about 150 km above the photosphere, and it spreads out (spatially) and gets 
weaker against height as seen from AIA 1700 and 1600 \AA~ emissions; this feature reflects the
spreading and weakening magnetic field (and hence increasing height for $\beta$ = 1 locations) against height.
iii) Frequencies of peak power gradually shift to higher values, along with a spreading in frequency extent, as height 
increases in the atmosphere; in the upper photosphere, power halos (from $v$, $I_{\rm{co}}$ and $I_{\rm{uv1}}$) 
exhibit twin peaks, one centered around 6 mHz and the other around 8 mHz. iv) On the whole, the largest power excess
is seen over horizontal magnetic-field locations (as inferred from the photospheric field) for each observable
(or height) and the largest among these are seen from the $I_{\rm{uv1}}$ 
at about 7.5 mHz. v) there are no significant changes in peak positions (in wavenumber $k_{\rm{h}}$) of ridges in power spectra
due to height-dependent changes in high-frequency power excess.

Overall, it is clear that the upper-photospheric and lower-chromospheric regions covered by the magnetic canopy
spatially redistribute incoming high-frequency acoustic-wave energy from below into a mixture of slow and fast 
magneto-acoustic waves, through mode conversions around the $\beta$ = 1 layer, so as to cause enhanced power around
photospheric strong fields. The $B$ and $\gamma$ dependences of power halos brought out in Figures 5 -- 11,
possibly include more intricate signatures of above wave interactions, which we have not been able to discern from
our current analyses. As mentioned earlier, magnetic-field extrapolations above the photospheric layers and models
of atmospheric structure need to be combined to estimate the height variation of $\beta$, and sound and Alfven speeds, which
in turn would lead to a clear and unambiguous identification of physical mechanisms behind the power halos. We intend to
follow up this present work with such attempts. 

\begin{acks}
S. Couvidat, K. Hayashi, and Xudong Sun are supported by NASA grants NNG05GH14G to the SDO/HMI project
at Stanford University. The data used here are courtesy of NASA/SDO and the HMI and AIA science teams. Data intensive computations 
performed in this work made use of the High Performance Computational facility of the Indian Institute of Astrophysics, Bangalore. Discussions with 
C.R. Sangeetha (IIA, Bangalore) are acknowledged. 
\end{acks}

\end{article} 

\end{document}